\documentclass[12pt]{article}
\usepackage{amsmath,color,epsfig,amsfonts,fullpage,natbib, amsthm, amssymb}
\usepackage[colorlinks=true,pdfstartview=FitV,linkcolor=blue,citecolor=blue,urlcolor=blue]{hyperref}
\usepackage{enumitem}
\usepackage{subcaption}
\usepackage{cleveref}
\usepackage{soul}
\usepackage[normalem]{ulem}
\usepackage{courier}
\newtheorem{theorem}{Theorem}[section]
\newtheorem{definition}[theorem]{Definition}
\newtheorem{lemma}[theorem]{Lemma}

\newcommand{\intp}{\int_{-\pi}^{\pi}}
\newcommand{\bd}{\boldsymbol}

\newcommand{\bbeta}{{\bd \beta}}

\newcommand{\vv}[1]{\mbox{\boldmath $#1$}}

\newcommand{\beq}{\begin{sxeqnarray*}}\newcommand{\eeq}{\end{eqnarray*}}
\newcommand{\beqn}{\begin{eqnarray}}\newcommand{\eeqn}{\end{eqnarray}}
\newcommand{\eit}{\end{itemize}}
\newcommand{\bnum}{\begin{enumerate}}\newcommand{\enum}{\end{enumerate}}
\newcommand{\nch}{\left\{\begin{array}{ll}}\newcommand{\ech}{\end{array}\right.}
\theoremstyle{definition}

\newtheorem{illus}{Illustration}[section]

\newcommand{\logit}{\textrm{logit}}
\newcommand{\invlogit}{\textrm{invlogit}}

\begin{document}

\begin{center}
\section*{Modeling Nonstationary Time Series using Locally Stationary
  Basis Processes}

\vspace{-.3cm}

\textbf{Shreyan Ganguly}$^{1,2}$ and
\textbf{Peter F. Craigmile}$^{1,3}$
\vspace{-.2cm}

$^1$
Department of Statistics,
The Ohio State University,
Columbus, OH 43210, USA
\vspace{-.2cm}

$^2$\verb_ganguly.28@osu.edu_ \;
$^3$\verb_pfc@stat.osu.edu_ \;

\vspace{-.2cm}
\textit{Last updated April 26, 2021}
\vspace{-.4cm}
\end{center}

\begin{abstract}
  Methods of estimation and forecasting for stationary models are well
  known in classical time series analysis.  However, stationarity is
  an idealization which, in practice, can at best hold as an
  approximation, but for many time series may be an unrealistic
  assumption.  We define a class of locally stationary processes which
  can lead to more accurate uncertainty quantification over making an
  invalid assumption of stationarity.  This class of processes assumes
  the model parameters to be time-varying and parameterizes them in
  terms of a transformation of basis functions that ensures that the
  processes are locally stationary.  We develop methods and theory for
  parameter estimation in this class of models, and propose a test
  that allow us to examine certain departures from stationarity.  We
  assess our methods using simulation studies and apply these
  techniques to the analysis of an electroencephalogram time series.

\textbf{Keywords}:
 Time varying processes; Tests of stationarity; Causality; Parameter estimation; Uncertainty quantification; EEG
\end{abstract}

\section{Introduction}\label{sec:Intro}

Weak or second order stationarity of a stochastic process, often after
detrending or deasonalization, is a pivotal assumption in the modeling
and analysis of time series.
There is an extensive range of statistical methods available for the
model selection, estimation, and forecasting of stationary time
series~\cite[see,
e.g.,][]{brillinger2001time,brockwell1991time,shumway2006time}.
However, even after accounting for possible trends or seasonality, the
assumption of stationarity can be unlikely, leading to compromised
inference of the stochastic processes of interest.  Areas of
application that lead naturally to nonstationary time series models
include, for example, acoustics~\cite[e.g.][]{eom1999analysis,
  averbuch2009wavelet}, biomedical
science~\cite[e.g.][]{prado2002time, dahal2014tvar}, climate
science~\cite[e.g.][]{takanami1991estimation}, and
oceanography~\cite[e.g.][]{whitcher2000wavelet}.

A flexible class of nonstationary models called \textit{locally
  stationary processes}~\citep{dahlhaus1997fitting} have gained in
popularity, due to their ability to model evolving time series
dependence, while allowing for efficient statistical estimation.  (See
\citet{priestley1965evolutionary} for the origins of evolving time
series dependence and \citet{dahlhaus2012locally} for a summary of
methods of inference for locally stationary processes.)  Traditional
representations of locally stationary models use time varying spectral
or infinite moving average
representations~\citep{dahlhaus1996kullback, dahlhaus1996maximum,
  dahlhaus2012locally}, but later extensions involve the use of local
exponential bases~\citep{Ombao:2001:Automatic, ombao2005slex} or
wavelet representations~\citep{neumann1997wavelet,nason2000wavelet}.
There are also multivariate
extensions~\cite[e.g.,][]{dahlhaus2000likelihood, park2014estimating,
  cardinali2017locally}.  While there is an extensive literature on
rigorous theoretical results for locally stationary
processes~\cite[e.g.][]{dahlhaus2012locally} and tests for
stationarity~\cite[e.g.][]{von2000wavelet,sakiyama2004discriminant,paparoditis2010validating,dwivedi2011test},
there is relatively less literature that demonstrate the
practicalities of fitting locally stationary models to actual data
(but see, e.g., \citet{palma2010efficient},
\citet{dahlhaus2012locally} and \citet{palma2013estimation}).

We introduce a class of locally stationary processes, called
\textit{locally stationary basis (LSB) processes} that are
parameterized by a transformation of basis functions that are able to
capture smooth changes in the time-varying time series parameters.
Using transformations of basis functions ensure that our models are
locally stationary, and as appropriate, causal, invertible or
identifiable.  We provide examples of LSB processes that are
nonstationary extensions of popular classes of stationary processes.
This includes introducing LSB processes that are related to the
popularly used time-varying autoregressive processes~\cite[e.g.][]{prado2002time, rudoy2011time}, as well as time-varying fractionally differenced
processes~\cite[e.g.][]{whitcher2000wavelet,palma2010efficient,Roueff:rvs:2010} that allow for long range
dependence~\cite[e.g.][]{beran:1994} that varies through time.
We demonstrate that the LSB processes facilitate practically viable
statistical inference: inference for LSB processes can be
statistically and computationally efficient, model selection and
forecasting follows naturally, and we can also test for departures
from stationarity in a straightforward manner.

In Section~\ref{sec:model}, we introduce LSB processes, define their
statistical properties, and provide a wide range of example processes
belonging to this class of models.
Parameter estimation, and associated asymptotic theory, along with
model selection and forecasting methodology is discussed in
Section~\ref{sec:estimation}.
Section~\ref{sec:test} introduces our test for stationarity
using this class of models.
Simulation studies to investigate parameter estimation and the
performance of the test for stationarity is shown in
Section~\ref{sec:simulation}.
We demonstrate the analysis of an EEG series as a practical
application in Section~\ref{sec:examples},
and close with some discussion in Section~\ref{sec:conc}.
Theoretical proofs and details of algorithms used in the
article are provided in Supplementary Material.

\section{Locally stationary basis processes (LSB processes)}
\label{sec:model}

Before we introduce LSB processes, we review the definition of univariate
locally stationary processes, taken from \citet{dahlhaus1997fitting}.

\begin{definition}\label{defn:local:stat}
  For a positive integer $T$, %
  $ \{X_{t,T} : t=1,\ldots,T \}$ belongs to the class of
  \textbf{locally stationary processes} with transfer function $ A^0 $
  and trend $ \mu $ if it has the spectral representation
  \begin{equation}\label{spec_rep}
    X_{t,T} = \mu\!\left(\frac{t}{T} \right) + \int_{-\pi}^{\pi} \exp(i\lambda t) A_{t,T}^0(\lambda) dZ(\lambda).
  \end{equation}
  where

(i)
  $\{ Z(\lambda) : \lambda \in [-\pi, \pi] \}$ is a stochastic process
  that satisfies $\overline{ Z(\lambda) } = Z(-\lambda)$ (here
  $\overline{Z}$ is the complex conjugate of $Z$) and has $k$th order
  cumulant,
\begin{equation*}
  \textrm{cum} \{d Z(\lambda_1) , \dots, d Z(\lambda_k) \} =
  \eta\!\left( \textstyle \sum_{j=1}^k \lambda_k \right) \;
  \gamma_k(\lambda_1, \dots, \lambda_{k-1} ) \;
  d\lambda_1, \dots, d\lambda_k,
\end{equation*}
with $\gamma_1 = 0$, $\gamma_2(\lambda)=1$, and
$|\gamma_k(\lambda_1, \dots, \lambda_{k-1} )|$ is bounded for all
$k$.  The function $\eta(\lambda) = \sum_{j=-\infty}^{\infty}
\delta(\lambda + 2\pi j)$ is the $2\pi$ extension of the Dirac delta function;

(ii) There exists a positive constant $K$ and a $2\pi$-periodic time
varying transfer function
$A:[0,1] \times \mathbb{R} \rightarrow \mathbb{C} $ with
$\overline{A(u, \lambda)} = A(u, -\lambda)$ and
  \begin{equation*}\label{lscond}
  \sup_{t,\lambda} \left |{A^0}_{t,T}(\lambda)-A\!\left(\frac{t}{T},\lambda \right)\right | \leq KT^{-1},
  \end{equation*} 
  for all $T$.  The functions $ A(u, \lambda) $ and $ \mu(u) $ are
  assumed to be continuous in $u$.
\end{definition}

In the above definition, $u = t/T$ defines the \textit{rescaled time}
unit.  This definition of rescaled time leads to infill asymptotic-based theory
for the study of locally stationary processes. When
$\{ Z(\lambda) \}$ is Brownian motion, the locally stationary
process is Gaussian.

We now define a rich class of locally stationary processes by
expressing the continuous time varying transfer function
$A(u, \lambda)$ in terms of smooth time varying parameter curves.

\begin{definition}\label{lsb_model}
  A \textbf{locally stationary basis (LSB) process} $ \{{X_{t,T}}\}$
  is a locally stationary process as given by Definition
  \ref{defn:local:stat} where the time varying transfer function
  $A(u, \lambda)$ is defined in terms of a continuous function of $J$
  time varying parameter curves
  \begin{eqnarray*}
  \{ \theta_j(u) : u \in [0,1] \}, \quad j=1,\ldots,J.
\end{eqnarray*}
Each parameter curve $\theta_j(u)$ is defined via
generalized linear functions of basis vectors:
\begin{eqnarray}\label{basisrep}
  g_j(\theta_j(u)) &=& \vv{w}_{j}'(u) \vv{\beta}_{j},  \quad u \in [0,1].    
\end{eqnarray}
For each $j=1,\ldots,J$, $g_j$ is a continuous and
differentiable 1-1 link function,
$\{ \vv{w}_j(u) \}$ is a vector of $b_j$ smooth basis functions, and $\vv{\beta}_j$
denotes a $b_j$-vector of model coefficients.
\end{definition}

Let $\vv{w} = \{ \vv{w}_1, \ldots, \vv{w}_J \}$ denote the entire
collection of $b = \sum_{j=1}^J b_j$ basis functions and $\vv{\beta} = (\vv{\beta}_1', \dots, \vv{\beta}_J')' = (\beta_{jl} : j=1,\ldots,J, l=0,..,b_j )'$ denote the complete set of model coefficients.  We suppose that $ \bbeta \in \mathcal{B}$, a closed subset of $\mathbb{R}^b$.  Often we will write $A(u, \lambda; \bbeta)$ to
emphasize the relationship between $\bbeta$ and the time varying transfer
function.
Modeling the time varying parameters with a transformation of basis
functions has appeal due to its flexibility in the choice of basis functions. The local stationarity of the process is
preserved because the time varying transfer function
$ A(u, \lambda; \bbeta) $ varies smoothly over rescaled time, $u$, and we
have a smooth function of well-behaved transformations of linear
combinations of smooth basis functions that guarantee that the time varying transfer function corresponds to the transfer function of a stationary process for each $u$ (see \citet{palma2013estimation} for an example of defining LS processes without the use of link functions). 

In practice, while it seems challenging at first to define a process
using time varying transfer functions, we will demonstrate in Section
\ref{sec:LSB:examples} that there are many examples of such transfer
functions available to us.  In
addition, a wide class of basis functions, such as Fourier,
polynomials, splines and wavelets yield flexible classes of LS processes.  Appropriate basis functions can be chosen
according to the problem at hand.

\subsection{Statistical properties of LSB processes}

By definition, LSB processes inherit the statistical properties of
LS processes.  For example, since $E(dZ(\lambda)) = 0$ for all $ \lambda$,
by definition of the LS process, we have that an LSB
process $\{ X_{t,T} \}$, following Definition \ref{lsb_model}, satisfies
$
    E(X_{t,T}) = \mu(t/T),
$
for all $t$ and $T$.
To define the time varying covariance function, we first define the
time varying spectral density function (SDF) $f(u, \lambda; \bbeta)$ by
\begin{eqnarray}\label{spectrum}
  f(u, \lambda; \bbeta)
  &=&
  | A(u, \lambda; \bbeta) |^2; 
\end{eqnarray}
that is, the time varying SDF is the modulus squared of the
time varying transfer function.  As \citet{dahlhaus2012locally}
explains for locally stationary processes, this time varying SDF
can be interpreted in terms of the Wigner-Ville spectrum which is
popularly used for time-frequency
analysis~\citep[e.g.][]{martin:1985:wigner,flandrin:1998:time}.  The
Wigner-Ville spectrum, $f_T(u, \lambda)$, is defined by local (in time)
Fourier transforms:
\begin{eqnarray*}
f_T(u, \lambda)
&=&
\frac{1}{2\pi} \sum_{h=-\infty}^{\infty} \text{cov}\left(X_{[uT-h/2],T}, X_{[uT+h/2],T}\right) \exp(-i\lambda h),
\end{eqnarray*}
where $X_{t,T}$ is set to zero for $t<1$ and $t>T$.
\citet{dahlhaus2012locally} shows that if we can represent $\{ X_{t,T}
\}$ as an LS linear process,
\begin{eqnarray}\label{eq:locally:linear}
     X_{t,T}
  &=&
      \mu(t/T) + \sum_{j=-\infty}^{\infty} \psi_{t,T,j} \epsilon_{t-j},
\end{eqnarray}
where $\mu$ is of bounded variation and $\{ \psi_{t,T,j} \}$ satisfies a
number of conditions given by \citet[][Assumption
4.1]{dahlhaus2012locally}, then the two time varying spectra are
related in the following sense:  for all $u \in [0,1]$,
\begin{align*}
\intp \left| f_T(u, \lambda) - f(u, \lambda; \bbeta) \right|^2 d\lambda = o(1).
\end{align*}
(Many of the LSB example processes in
Section~\ref{sec:LSB:examples} can be written in the form given by
\eqref{eq:locally:linear}.)  We then define the time varying
covariance function for our LSB process, $c(u, h; \bbeta)$, at
rescaled time $u$ and lag $h$ by
\begin{eqnarray}\label{eq:lscov}
    c(u, h; \bbeta) = \intp f(u, \lambda; \bbeta) \exp(ih\lambda) d\lambda.
\end{eqnarray}
If we define $ \text{cov}(X_k,X_l) $ as $c^0(k,l) $, then an additional restriction on $\{ \psi_{t,T,j} \}$,
\citet[][Equation 69]{dahlhaus2012locally} shows that uniformly in u and h,
\begin{equation}\label{eq:tvcov}
\text{cov} \left(X_{[uT-h/2],T}, X_{[uT+h/2],T}\right) = c^0(uT-h/2,uT+h/2) = c(u, h; \bbeta) + O\left(T^{-1}\right).
\end{equation}

\subsection{Example LSB processes}\label{sec:LSB:examples}

\textbf{A simple example:}
Before we introduce general classes of LSB processes, we start with a simple
example.  Suppose that $\{ \epsilon_t \}$ is an independent Gaussian
process with mean zero and variance 1.  This stationary process has the following
spectral representation
\begin{equation*}
\epsilon_t =  \intp \exp(i\lambda t) \; A(\lambda) \; dZ(\lambda),
\end{equation*}
where the (constant-in-time) transfer function $A(\lambda) = 1/\sqrt{2\pi}$ for
all $\lambda$ and $\{ Z(\lambda) \}$ is a Brownian motion.  Now
consider a variance-modulated process $\{ X_{t,T} \}$ that rescales
the process $\{\epsilon_t\}$ by a smoothly varying standard deviation
(SD) function $\sigma(t/T)$:
\begin{eqnarray*}
    X_{t,T} = \sigma(t/T) \epsilon_t, \quad t=1,\ldots,T.    
\end{eqnarray*}
We model the SD curve $\{ \sigma(u) : u \in [0,1] \}$ on the log
scale to preserve positivity of variances.   Using the log link
function $g(x) = \log(x)$ and a set of basis functions $\vv{w}(u)$
to model the log SD curve, suppose that
\begin{eqnarray*}
    g(\sigma(u)) = \log \sigma(u) = \vv{w}'(u) \bbeta,    
\end{eqnarray*}
for some model parameters $\bbeta$.  To further simplify our example,
suppose that the log SD curve is a linear function of $u$: let
$\vv{w} = (1,u)'$ and $\bbeta = (\beta_0, \beta_1)'$.
The process $\{ X_{t,T} \}$ is an example of an LSB process:  our
spectral representation is
\begin{eqnarray*}
    X_{t,T} = \intp \exp(i \lambda t) \; A(t/T,\lambda; \bbeta) \; dZ(\lambda),
\end{eqnarray*}
where the transfer function $A(u, \lambda; \bbeta)$ is given by
\begin{eqnarray*}
A(u, \lambda; \bbeta) = \frac{1}{\sqrt{2\pi}}\sigma(u) = \frac{1}{\sqrt{2\pi}}\exp( \beta_0 + \beta_1 u),
\end{eqnarray*}
for each $u$ and $\lambda$.  Locally in time this process is a white
noise process (at each $u$ the transfer function is constant over
$\lambda$).  The time-varying SDF,
$f(u, \lambda; \bbeta) = |A(u, \lambda; \bbeta)|^2 = \exp( 2 \beta_0 +
2 \beta_1 u)/2\pi$ is constant over $\lambda$ but varies over rescaled time $u$. The time-varying covariance function,
\begin{eqnarray*}
  c(u, h; \bbeta)
&=&
    \intp f(u, \lambda; \bbeta) \exp(ih\lambda) d\lambda
\;=\;
 \begin{cases}
  \frac{1}{2\pi} \exp( 2 \beta_0 + 2 \beta_1 u), & h = 0 \\
  \ 0, & h \neq 0, 
  \end{cases}
\end{eqnarray*}
also varies over rescaled time u.  In
this simple example the time-varying transfer function has a simple
functional form. We now demonstrate more involved examples of time varying transfer functions.


\textbf{LSB-AR($p$) processes:}
Time varying autoregressive (AR) processes are the most commonly used
nonstationary process for time series analysis.
For a locally stationary AR process of order $p$, let
$\{ \phi_{p,j}(u) \}$ denote the $j$th AR parameter curve
($j=1,\ldots,p$), $\{ \sigma(u) \}$ denote the time-varying scale
curve, and suppose that $\{ \epsilon_{t,T} \}$ is a stochastic process with mean
0 and variance $ \sigma(t/T) $.  Then the process $\{ X_{t,T} \}$ is the solution to
\begin{eqnarray}\label{eq:TVAR}
    X_{t,T} = \sum_{j=1}^p \phi_{p,j}(t/T) X_{t-1, T} + \epsilon_{t,T}.
\end{eqnarray}
In practice, it can be difficult to ensure that the time varying autoregressive
parameter curves lead to locally stationary models. As in the
stationary case, this is related to the roots of the AR polynomial $\phi$
which, in addition to being a function of $z \in \mathbb{C}$, also depends on the local time point $u$:
\begin{eqnarray*}
    \phi(z, u) = 1 - \sum_{j=1}^p \phi_{p,j}(u) z^j.
\end{eqnarray*}
We also need continuity of the parameter curves over $u$.
\citet{kunsch1995note} provides the following.

\begin{definition}
The process defined by  \eqref{eq:TVAR} has the causal solution 
\begin{equation*}
  X_{t,T} = \sum_{k=0}^\infty \psi_{t,T,k} \epsilon_{t-k,T}
\quad\mbox{with}\quad
  \sup_{t,T} \sum_{k=0}^\infty |\psi_{t,T,k}| < \infty,
\end{equation*}
if $ \phi_{p,j}(u) $ is continuous on $[0,1]$ for all $j = 1, \dots, p$
and there exists a $ \delta > 0 $ such that $\phi(z, u) \neq 0$ for
all $ |z| \leq 1+\delta $ and for all $u$.
\end{definition}

When the process is causal it follows from standard linearity
filtering methods~\cite[e.g.][Chapter 5]{perc:wald:1993} that the transfer function is
\begin{eqnarray}\label{eq:ar_sdf}
  A(u, \lambda)
  \;=\;
  \frac{\sigma(u)}{\sqrt{2\pi}} \;
  \phi(\exp(-i\lambda), u)
  &=& \frac{\sigma(u)}{\sqrt{2\pi}}
      \left(1-\sum_{j=1}^p \phi_{p,j}(u)\exp(-ij\lambda) \right)^{-1}.
\end{eqnarray}

Instead of modeling the $\{ \phi_{p,j}(u) \}$ parameter curves
directly, we model the time varying partial autocorrelation curves.
In the stationary case, \citet{jones1980maximum} shows that for
AR($p$) processes there is a casual solution if and only if all the
$p$ partial autocorrelation parameters lie in $(-1, 1)$.  In the
time-varying case, let $\{ \phi_{j,j}(u) \}$ denote the partial
autocorrelation parameter curves which can be defined recursively and
efficiently via the Levinson-Durbin (LD) algorithm~\cite[e.g.,][]{brockwell1991time} at each local
rescaled time point $u$.

 For local
stationarity, we then require that each partial autocorrelation curve is
continuous in $u$ and takes values in the region $(-1,1)$.  To achieve
this we complete the definition of the LSB-AR($p$) process as follows.
Let $g(x) = \logit(x) = \log(x/(1-x))$ denote the logit function.
Then for each $j=1,\ldots,p$, suppose that the partial autocorrelation
parameter curves satisfy
\begin{eqnarray*}\label{ar_cond}
  g( (\phi_{j,j}(u) + 1)/2 )
  &=&
      \vv{w}_j'(u) \; \bbeta_j,
\end{eqnarray*}
for $p$ sets of basis functions $\{ \vv{w}_j(u) \}$
($j=1,\ldots,p$).  Using a log link function and another set of basis
functions $\{ \vv{w}_{p+1}(u) \}$, we model the scale curve using
%
$
\log \sigma(u)
\;=\;
      \vv{w}_{p+1}'(u) \; \bbeta_{p+1}.
      $

\begin{figure}[t]
\begin{center}
\includegraphics[scale = 0.8]{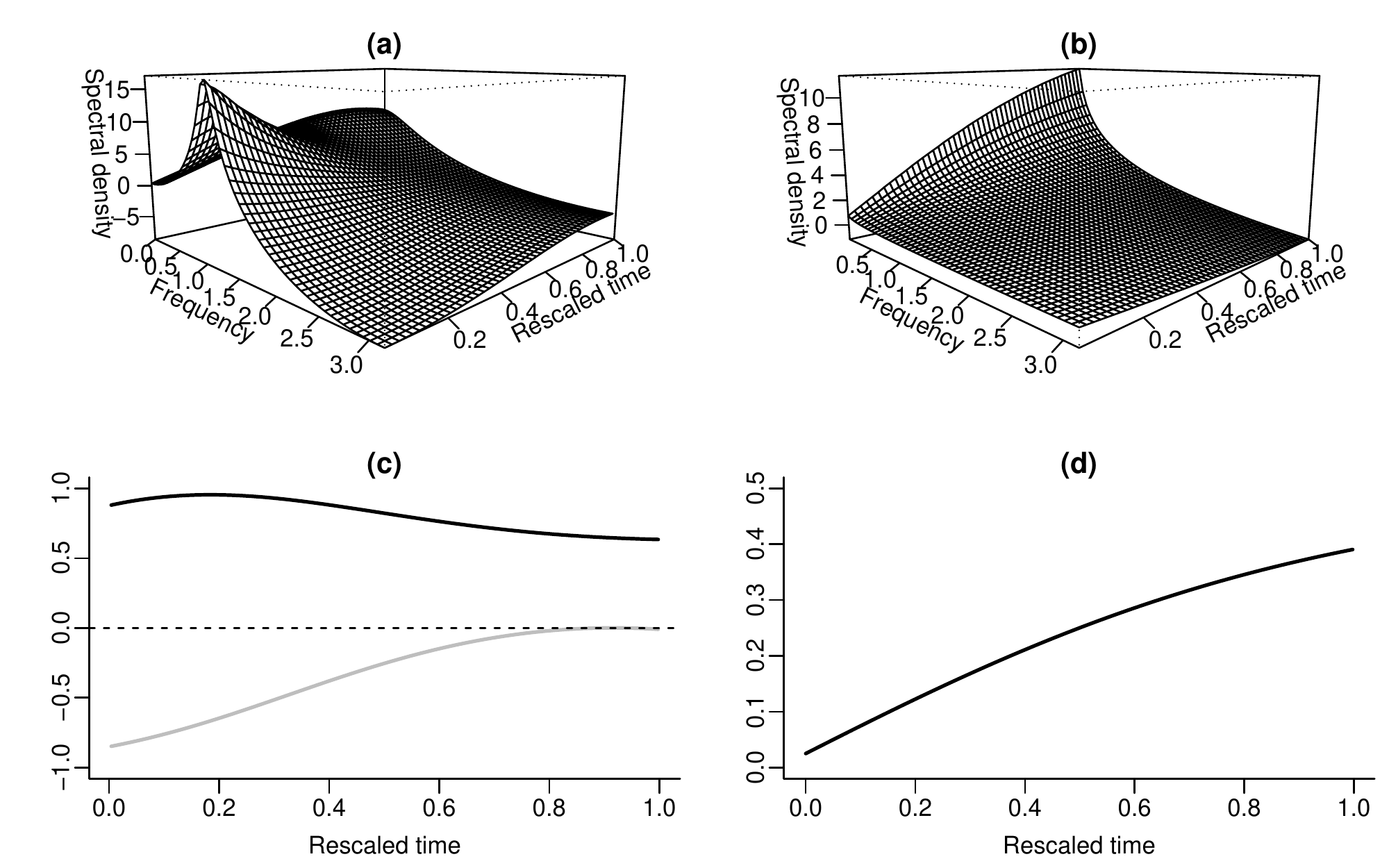}
\caption{Panels (a) and (b) show the plot of the spectral density of an LSB-AR process of order 2 and an LSB-FD process. Panel (c) shows the time varying AR curves of order 1 (black) and 2 (gray) that characterizes the LSB-AR process, while panel (d) shows the time varying LRD curve that characterizes the LSB-FD process. }
\label{AR2_FD}
\end{center} 
\end{figure}

Consider the interesting example of an LSB-AR process of order 2 that slowly evolves into an LSB-AR process of order 1.  We let
\begin{eqnarray*}
	g( (\phi_{j,j}(u) + 1)/2 )
	&=&
	\beta_{j0} + \beta_{j1} u + \beta_{j2} (u^2 - 1/3), \quad j = 1,2,
\end{eqnarray*}
with $\beta_{10}=0.61$, $\beta_{11}=1.71$, $\beta_{12} = -1.27$, $\beta_{20}=-3.52$, $\beta_{21}=5.50$, $\beta_{22} = -3.00$ and
$\sigma(u) = 1$ for all $u$. The time-varying SDF $f(u, \lambda;\bbeta) = |A(u, \lambda; \bbeta)|^2$, with $ A(u, \lambda; \bbeta) $ given by \eqref{eq:ar_sdf} is shown in panel (a) of Figure~\ref{AR2_FD}. The time varying AR parameter curves $ \phi_{2,1}(u) $ (in black) and $ \phi_{2,2}(u) $ (in gray), calculated using the LD, are displayed in panel (c) of Figure~\ref{AR2_FD}. The  time-varying SDF shows the peak at a non-zero frequency in the AR(2) SDF slowly disappearing as $ \phi_{2,2}(u) $ goes to 0 as u approaches 1.


Adding time-varying moving average components leads
naturally to the LSB-MA($q$) and LSB-ARMA($p$, $q$) processes. 


\textbf{LSB-Exp($p$) processes:}
The exponential (Exp) model of \citet{bloomfield1973exponential} is a
popular stationary time series process, especially when log SDFs are
estimated via regression
models~\cite[e.g.][]{wahba1980automatic,moulines1999broadband}.  The
stationary model represents the log SDF via a truncated Fourier
series, and provides for a simple way to estimate the parameters using
spectral estimates of the data.
It is natural to consider
the case where the SDF varies over time; see, e.g.,~\citet{Rosen_2009} for a
Bayesian mixture version of this idea.  An LSB-Exp
process of order $p$ has time-varying transfer function
\begin{equation*}
A(u, \lambda) = \frac{\sigma(u)}{\sqrt{2\pi}}\sqrt{\exp \left
    \{2\sum_{j=1}^p \theta_j(u) \cos(j\lambda) \right \}},
\end{equation*}
where $\{ \theta_j(u) \}$ ($j=1,\ldots,p$) are $p$ different parameter
curves and $\{ \sigma(u) \}$ is the time-varying scale parameter curve.
In the stationary case, $\theta_j$ for $j=1,\ldots,p$ (without $u$)
are known as cepstral coefficients~\citep{bogert1963quefrency}, and so
we can refer to $\{\theta_j(u)\}$ as \textit{time-varying
cepstral curves}.
These models are locally stationary if all the cepstral curves and
scale parameters are continuous in $u$ and positive.   Letting $g(x) =
\log(x)$
denote the log link function, we complete our LSB-EXP process by supposing
\begin{eqnarray*}
    g( \theta_j(u) ) &=& \vv{w}_j(u)' \bbeta_j, \qquad j=1,\ldots,p; \\
    g( \sigma(u) ) &=& \vv{w}_{p+1}(u)' \bbeta_{p+1},
\end{eqnarray*}
for $p+1$ basis functions $\{ \vv{w}_j(u) \}$, $j=1,\ldots,p+1$.


\textbf{LSB-FD processes:}
Stationary long range dependent (LRD) processes, also known as long memory
processes, are characterized by a correlation function that slowly
decays to zero.  Equivalently, the SDF of the process has
a pole at zero frequency. (See \cite{beran:1994} for a review of
statistical methods for LRD processes.)  While there are
definitions of nonstationary LRD processes constructed via random
walks~\cite[e.g.][]{gra:joy:1980,hosking:1981}, one can also define
locally stationary LRD processes.

As simple example, consider the time-varying fractionally differenced
(FD) process~\citep{whitcher2000wavelet,palma2010efficient,Roueff:rvs:2010}.  Using a
basis representation, we can define LSB-FD processes as follows.
Define the time-varying transfer function  $A$ for $\{ X_{t,T} \}$, via,
\begin{eqnarray}\label{eq:transfer:LSBFD}
  A(u, \lambda)
  &=&
 \frac{\sigma(u)}{\sqrt{2\pi}} \left\{1 - \exp(-i\lambda)\right\}^{-\delta(u)},
\end{eqnarray}
where $\{ \delta(u) : u \in [0,1] \}$ is the time-varying LRD
parameter curve, and $\{ \sigma(u) : u \in [0,1] \}$ is the
time-varying scale parameter curve.  The process $\{X_{t,T}\}$ is locally
stationary if $\{ \delta(u) \}$ takes values on $(-1/2, 1/2)$,
$\{ \sigma(u) \}$ takes positive values, and both curves are
continuous in $u$~\citep{whitcher2000wavelet,Roueff:rvs:2010}.  Let
$\{ \vv{w}_1(u) \}$ denote a set of basis functions for
$\{ \delta(u) \}$.  Then we define
\begin{eqnarray*}\label{eq:fd1}
    g_1(\delta(u)) = \vv{w}_1'(u) \bbeta_1,
\end{eqnarray*}
for the link function $g_1(x) = \logit(x + 1/2)$.  For another set of basis function
$\{ \vv{w}_2(u) \}$ for the SD curve we let
%
$ g_2
    (\sigma(u)) = \vv{w}_2'(u) \bbeta_2, $
with the link function $g_2(x) = \log x$.

Panel (b) of Figure~\ref{AR2_FD} presents the the time-varying SDF
$f(u, \lambda; \bbeta) = |A(u, \lambda; \bbeta)|^2$, with $A(u, \lambda; \bbeta) $  given by \eqref{eq:transfer:LSBFD}, of an LSB-FD process with
$g_1(\delta(u)) = \beta_0 + \beta_1 u$, with $\beta_0=0.1$, $\beta_1=2.5$, and
$\sigma(u) = 1$ for all $u$.  The time-varying LRD parameter curve
$\delta(u)$ is shown in panel (d) of the same figure.  For all local
time points, $u$, this process has a pole at zero frequency, but given
the fact that $\beta_1$ is positive the process becomes more LRD as we
move from $u=0$ to $u=1$.  Note that the case of $\beta_1 = 0$
correspond to the process being a stationary FD process with
(non-time-varying) LRD
parameter $\delta = \invlogit(\beta_0) - 1/2$, where $\invlogit$ is
the inverse logit function.  We use this idea as a
more general basis for testing for stationarity in Section \ref{sec:test}.

As is common in the stationary case, we can extend
this model by including (locally) stationary ARMA or exponential
components to the transfer function \eqref{eq:transfer:LSBFD}.  This
allows us to capture time-varying short range dependence as well as
LRD.  For example, LSB autoregressive fractionally integrated moving
average (LSB-ARFIMA) processes have time varying transfer function
\begin{equation*}
A(u, \lambda) = \frac{\sigma(u)}{\sqrt{2\pi}} \frac{1+ \sum_{j=1}^q
  \theta_j(u)\exp(-ij\lambda)}{1 - \sum_{k=1}^p \phi_k(u)\exp(-ik\lambda)} \left\{1 - \exp(-i\lambda)\right\}^{-\delta(u)}.
\end{equation*}
Here, the AR and MA components of the process are parameterized using basis functions in a similar manner as the LSB-AR processes presented above.

\section{Statistical inference}
\label{sec:estimation}

Assume that the LSB process is correctly specified
and is Gaussian.  Without loss of generality, we also assume that the
process has mean zero.  Suppose we have a finite sample
$ \boldsymbol{X}_T = (X_{1,T}, \dots, X_{T,T})' $ of T observations
drawn from an LSB process $ \{X_{t,T}\} $ defined as in
Definition~\ref{lsb_model} with true time varying transfer function
$ A^0_{t,T}(\lambda; \bbeta_0) $, where the model parameters
$ \bbeta_0 \in \mathcal{B} $.
%
  %
We introduce likelihood and block Whittle likelihood estimators for
the model parameters in Sections~\ref{sec:exact} and \ref{sec:whittle}
respectively.  In Section~\ref{sec:large:sample} we provide large
sample properties of these estimators and we use these theoretical
results to provide inference for the parameter curves
$\{ \theta_j(u) \}$ in Section~\ref{sec:inference:curves}.  We discuss
model selection and forecasting for LSB processes in
Sections~\ref{sec:model:selection} and \ref{sec:forecasting}, respectively.

  \subsection{Likelihood-based estimation}
  \label{sec:exact}

The negative log-likelihood for $\boldsymbol \beta$
using data $\boldsymbol{X}_T$, normalized by the sample size $T$, is
\begin{align}\label{exactlike}
  \mathcal{L}_T(\boldsymbol \beta) 
=& \frac{1}{2} \log(2\pi) + \frac{1}{2T} \log \det \bd \Sigma_{\boldsymbol \beta} + \frac{1}{2T} (\bd X_T'\bd \Sigma_{\boldsymbol \beta}^{-1} \bd X_T).
\end{align}
  Here the model covariance matrix   is a function of the time varying transfer function
  $ A(u, \lambda; \bbeta) $:
  \begin{equation}\label{exactlikesigma}
  \bd \Sigma_\bbeta = \left\{c^0(k,l;\bbeta)\right\}_{k,l = 1, \dots, T}
  = \left\{ \intp\exp(i\lambda(k-l)
    \textstyle
    A\!\left(\frac{k}{T}, \lambda; \bbeta\right)
    A\!\left(\frac{l}{T},-\lambda; \bbeta\right) d\lambda \right\}_{k,l = 1, \dots, T}.
  \end{equation}
Then the maximum likelihood estimate $\widehat{\bbeta}_T $ of $
\boldsymbol \beta $ is
\begin{align*}
  \widehat{\bbeta}_T = \text{arg}\min_{\bbeta \in \mathcal{B}} \mathcal{L}_T(\bbeta).
\end{align*}
Since the calculation of the determinant and inverse of
$ \bd \Sigma_\bbeta $ is computationally intensive we use a
modified Cholesky decomposition approach to compute the likelihood \eqref{exactlike}.
Let
\begin{align}\label{eqBLP}
  \widehat{X}_{t,T} =&
                       \nch
                       0, & t=1; \\
  \sum_{k=1}^{t-1} \phi_{t-1,k}(u)
  X_{t-k,T}, & t = 2, \dots, T,
               \ech
\end{align}
denote the best linear predictor (BLP) of $ X_{t,T} $ given
$ \{ X_{1,T}, \dots, X_{t-1,T}\} $.
In \eqref{eqBLP}, the time dependent partial regression coefficients $
\phi_{t,k}(u) $ for $ t = 1, \dots T-1 $ are calculated using the LD algorithm and are given by
\begin{align}\label{ldrec}
\begin{split}
\phi_{t,t}(u) =& \ \left[c(u,t; \bbeta) - \sum_{j=1}^{t-1} \phi_{t-1,j}c(u,t-j; \bbeta)\right]/\sigma_{t,T}^2, \\
\phi_{t,k}(u) =& \ \phi_{t-1,k}(u) - \phi_{t,t}(u)\phi_{t-1,t-k}(u), \quad k = 1, \dots, T-1, 
\end{split}
\end{align}
where the prediction variances $\{ \sigma^2_{t,T} \}$ are given by
\begin{eqnarray*}
  \sigma_{t,T}^2 &=&
                     \nch
                     c^0(1,1; \bbeta), & t=1; \\
  \sigma_{t-1,T}^2 \left[1 - \phi_{t,t}^2(u)\right], & t = 2, \dots, T.
                     \ech
\end{eqnarray*}
In the above equations $ c(u, \cdot; \bbeta) $ comes from \eqref{eq:lscov} and 
$ c^0(\cdot, \cdot; \bbeta) $ is given \eqref{eq:tvcov}. 
Then, letting $ \epsilon_{t,T} = X_{t,T} - \widehat{X}_{t,T}$ be the partial
innovations with variance $ \sigma^{2}_{t,T} $, we rewrite \eqref{exactlike} as
\begin{equation}\label{predlik}
  \mathcal{L}_T(\bbeta) = \frac{1}{2}\log(2\pi) + \frac{1}{2T}\sum_{t=1}^T \left\{\log \sigma^{2}_{t,T} + \frac{{\epsilon^{2}_{t,T}}}{\sigma^{2}_{t,T}}\right\}. 
  \end{equation}
In practice we minimize \eqref{predlik} with respect to $\bbeta$ using the BFGS
numerical
solver~\citep[see][]{broyden1970convergence,fletcher1970new,goldfarb1970family,shanno1970conditioning}.
We discuss a matrix-version of this calculation in Section S1 of
  the Supplement.

The algorithmic complexity of implementing this time varying version
of the LD algorithm for any LS linear process is $ O(T^3) $. However,
for time varying Markov processes such as the LSB-AR process, this
algorithm can be implemented in $ O(T^2) $ operations (For an LSB-AR(p) model we have $ O(pT^2) $).
For non-Markov models such as LSB-FD processes, it is common to
approximate the likelihood by approximating $\widehat{X}_{t,T}$ using
a finite number, say $d$, of observations from the past. There is no
fixed method to choose $ d $ and the choice typically depends
on the degree of non-stationarity in the series.  A data adaptive
method for determining $ d $ appears in
\citet{fryzlewicz2003forecasting}.


\subsection{Block Whittle-based estimation}
\label{sec:whittle}

  Although we will show in Section~\ref{sec:large:sample} that the
  likelihood estimator $ \widehat{\bbeta}_T$ has desirable large
  sample properties such as consistency and asymptotic normality, the
  computational cost of this method can be high, especially for
  non-Markov processes such as long memory LSB processes. A standard
  alternative is to approximate the likelihood function
  using a nonstationary variation of the Whittle likelihood known as
  the Block-Whittle likelihood~\citep{dahlhaus1997fitting,
    palma2010efficient}, which is given by
 \begin{equation}\label{whitlike}
  \mathcal{L}_T^{W}(\bbeta) = \frac{1}{4\pi} \frac{1}{M} \sum_{j=1}^M \intp \left[\log 4\pi^2 f(u_j, \lambda; \bbeta) + \frac{I_N(u_j, \lambda)}{f(u_j, \lambda; \bbeta)} \right] d \lambda.
\end{equation}
Here $ f(u, \lambda; \bbeta) $ is defined by
(\ref{spectrum}) and $ I_N(u, \lambda)  $ is the local tapered periodogram over a segment of length $N$ with midpoint $ [uT] $ defined as
  $$ I_N(u, \lambda) = \frac{1}{2\pi} \left|\sum_{s=0}^{N-1} \tau\!\left(\frac{s}{N}\right) X_{[uT]-N/2+s+1,T} \exp(-i\lambda s)\right|^2, $$ 
  where $ \tau (\cdot) $ is a data taper with $ \tau(x) = 0 $ for $ x
  \notin (0,1] $ and $ \sum_{x=0}^{N-1} \tau^2(x) = 1$, $u_j = t_j/T$
  and $ t_j = S(j-1) + N/2 $ for $ j = 1, \dots,
  \left[1+(T-N)/S\right]$.
%
The data taper is applied to the local periodogram
  to reduce the bias due to nonstationarity on a segment -- without the taper, it is not
  possible to achieve $ \sqrt{T} $-consistency for the block Whittle
  likelihood estimator.
 The block Whittle likelihood estimate $ \widehat{\bbeta}_T^{W} $ of $ \bbeta $ is 
 then $$ \widehat{\bbeta}_T^{W}  = \text{arg}\min_{\bbeta \in \mathcal{B}}
  \mathcal{L}_T^W(\bbeta), $$
 and is solved numerically via the BFGS algorithm.


\subsection{Large sample theory}
\label{sec:large:sample}

Again suppose that $ \bd X_T$ is our series of length $T$, from an LSB
process $ \{X_{t,T}\} $ defined as in Definition~\ref{lsb_model} with
true time varying transfer function $ A^0_{t,T}(\lambda; \bbeta_0) $,
where the model parameters $ \bbeta_0 \in \mathcal{B} $.  We further
suppose that $\bbeta_0$ exists uniquely and lies in the interior of
$\mathcal{B} $. We also assumed at the beginning of the section that the process is correctly specified and is Gaussian. We will show that both the exact likelihood estimator
$ \widehat{\bbeta}_T$ and the block Whittle estimator
$\widehat{\bbeta}_T^{W}$ are consistent, asymptotically normal, and
efficient for $\bbeta_0$ under certain assumptions.

We provide two alternative sets of conditions, depending on whether or
not the LSB process is LRD.  We say that the LSB process is LRD when
the time-varying SDF
$f(u, \lambda; \bbeta) = |A(u, \lambda; \bbeta)|^2$ has a pole at zero
frequency, $\lambda=0$, for some $u \in [0,1]$.  When the LSB process
is not LRD, we say that the process is \textit{short range dependent
  (SRD)}.  For the SRD case we make the following assumptions about
the time-varying transfer function and SDF.

\bnum

\item[] (AS) The time varying transfer function
  $ A(u, \lambda; \bbeta) $  is differentiable in $ u $ and $ \lambda $
  with uniformly bounded derivatives.  The time varying SDF
  $ f(u, \lambda; \bbeta)$ is strictly positive and is uniformly
  bounded from above and below.  If $\nabla$ denotes a derivative operator such that $ \nabla f(u, \lambda; \bbeta) = (\nabla_1 f(u, \lambda; \bbeta), \dots, \nabla_b f(u, \lambda; \bbeta)' $ where $ \nabla_k f(u, \lambda; \bbeta) = \frac{\partial}{\partial \beta_k} f(u, \lambda; \bbeta) $ and $ \nabla^2 f(u, \lambda; \bbeta) = \left[ \nabla_{kl} f(u, \lambda; \bbeta) \right]_{k,l = 1, \dots, b} $ where $ \nabla_{kl} f(u, \lambda; \bbeta) = \frac{\partial^2}{\partial {\beta_k}{\partial \beta_l}} f(u, \lambda; \bbeta) $, then  
  $\nabla f(u, \lambda; \bbeta) $ and
  $ \nabla^2 f(u, \lambda; \bbeta) $ are continuous on
  $ \mathcal{B} $. Also, for the reciprocal SDF,
  $ \nabla f(u, \lambda; \bbeta)^{-1} $ and
  $ \nabla^2 f(u, \lambda; \bbeta_0)^{-1} $ are differentiable in
  $ u $ and $ \lambda $ with uniformly bounded derivative.
  
  \enum

  For the LRD case we replace (AS) by (AL).  In the definition of (AL), $\delta(u)$ is
  the time-varying LRD parameter.
  
\bnum

\item[] (AL) The time varying SDF $ f(u, \lambda; \bbeta) $ is strictly
  positive and satisfies 
  \begin{align*}
  f(u, \lambda; \bbeta) \sim C_f(u, \bbeta)|\lambda|^{-2\delta(u)} \ \text{as} \ |\lambda| \rightarrow 0,
  \end{align*}
  where
  $ C_f(u, \bbeta) > 0, -\frac{1}{2} < \inf_{\bbeta, u}\delta(u),
  \sup_{\bbeta, u} \delta(u) < \frac{1}{2}$  and $ \delta(u) $ has
  bounded first derivative with respect to u. There is an integrable function $ g(\lambda) $ such that \\
  $ |\nabla_\bbeta \log f(u, \lambda; \bbeta) | \leq g(\lambda) $ for
  all $ \bbeta \in \mathcal{B}, u \in [0,1]$ and
  $ \lambda \in [-\pi, \pi]. $ The function $ A(u, \lambda; \bbeta) $
  is twice differentiable with respect to $ u $ and satisfies
\begin{equation*}
\int_{-\pi}^{\pi} A(u, \lambda; \bbeta) A(v, -\lambda; \bbeta) \exp\{ik\lambda\} d\lambda \sim C(u, v, \bbeta) k^{\delta(u) + \delta(v) -1},
\end{equation*}
as $ k \rightarrow \infty $, where $ |C(u, v, \bbeta)| \leq \text{const}. $ for $ u, v \in [0,1] $ and $ \bbeta \in \mathcal{B}$. The SDF $ f(u, \lambda; \bbeta)^{-1} $ is twice differentiable over $ \bbeta, u$ and $ \lambda$. 
  
  \enum

Additionally for the block Whittle estimator, we need to
provide conditions on the data taper $\tau$, the block length $N$, and step
size $S$.

For the SRD case we have
\bnum

\item[] (AWS) The block length $N$, step size $S$ and sample size $T $
  fulfill $ T^{\frac{1}{4}} \ll N \ll T^{\frac{1}{2}}/\log T $
  and $ S/N \rightarrow 0 $.  Also, the data taper
  $ \tau:\mathbb{R} \rightarrow \mathbb{R}$ with $ \tau(x) = 0$ for all
  $ x \notin [0,1] $ is continuous on $ \mathbb{R} $ and twice
  differentiable at all $ x \notin P $ where $ P $ is a finite set and
  $ \sup_{x \notin P} |\tau''(x)| < \infty $.

  \enum

 For the LRD case:

\bnum

\item[] (AWL) The block length $N$, step size $S$, block size $M$, and
  sample size $T$ satisfy
  $ S/N \rightarrow 0, \sqrt{T} \log^2N/N \rightarrow 0, \sqrt{T}/M
  \rightarrow 0 $ and $ N^3\log^2N/T^2 \rightarrow 0$.
The data taper $ \tau(x) $ is a positive, bounded function for $ x \in [0,1] $ and symmetric around $1/2$ with a bounded derivative. 
  \enum

  Then,  the following theorems hold for the exact likelihood
  estimator $\widehat{\bbeta}_T$ under (AS) for LSB-SRD and (AL) for LSB-LRD processes, and for the block Whittle estimator $\widehat{\bbeta}_T^W $ under additional assumptions (AWS) for LSB-SRD processes and (AWL) for LSB-LRD processes. 
  
\begin{theorem}[Consistency]\label{consis}
Both $ \widehat{\bbeta}_T \xrightarrow{P} \bbeta_0$  and $ \widehat{\bbeta}_T^{W} \xrightarrow{P} \bbeta_0$ as $ T \rightarrow \infty $. 
\end{theorem}

\begin{theorem}[Asymptotic Normality]\label{asy normal}
For the case when $ \mu(u) = 0$, 
\begin{equation*}
\sqrt{T}\left(\widehat{\bbeta}_T - \bbeta_0 \right) \xrightarrow{\mathcal{L}} \mathcal{N}\left(0, \bd \Gamma_{\bbeta_0}^{-1} \right) \ \text{and} \ \sqrt{T}\left(\widehat{\bbeta}_T^{W} - \bbeta_0 \right) \xrightarrow{\mathcal{L}} \mathcal{N}\left(0, \bd \Gamma_{\bbeta_0}^{-1} \right) \ \text{as} \ T \rightarrow \infty 
\end{equation*} 
where the Fisher information matrix for $ \bbeta $ is
\begin{equation}\label{inf_mat}
\bd \Gamma(\bbeta) =  \frac{1}{4\pi} \int_0^1 \intp (\nabla \log f(u,\lambda; \bbeta))(\nabla \log f(u, \lambda); \bbeta)' d\lambda du.
\end{equation}
\end{theorem}

\begin{theorem}[Efficiency]\label{efficiency}
Both the exact ML estimate $\widehat{\bbeta}_T$ and the approximate Whittle estimate $\widehat{\bbeta}_T^{W}$ are asymptotically Fisher efficient. 
\end{theorem}

 The Fisher information matrix $ \Gamma(\bbeta) $ given
 in~\eqref{inf_mat} is straightforward to evaluate. Let
 \begin{equation*}
   \nabla_{\bd j} \log f(u, \lambda; \bbeta)
   = \left(\frac{\partial}{\partial \bbeta_{j1}} \log f(u, \lambda; \bbeta)', \dots, \frac{\partial}{\partial \bbeta_{j b_j}} \log f(u, \lambda; \bbeta)'\right)'
 \end{equation*}
for $ j = 1, \dots, J$. 
Letting $ \eta_j(u) = g(\theta_j(u)), j = 1, \dots, J $, we have 
\begin{align*}
\frac{\partial}{\partial \beta_{jl}} \log f(u,\lambda; \bbeta) =& \frac{\partial}{\partial \theta_j(u)} \log f(u, \lambda; \theta_j(u)) \times \frac{\partial}{\partial \eta_j(u)} \theta_j(u) \times \frac{\partial}{\partial \beta_{jl}} \eta_j(u) \\
 =& \frac{1}{f(u,\lambda; \bbeta)} \frac{\partial}{\partial \theta_j(u)} f(u, \lambda; \theta_j(u)) \times \frac{1}{g'(\theta_j(u))} \times w_l(u).
\end{align*}
Thus, $ \bd \Gamma(\bbeta) $ is a $b\times b$ block diagonal matrix with submatrices ($j = 1, \dots, J$)
\begin{align*}
\bd \Gamma(\bbeta_j) =& \frac{1}{4\pi} \int_0^1 \intp (\nabla_{\bd j} \log f(u,\lambda; \bbeta))(\nabla_{\bd j} \log f(u, \lambda; \bbeta))' d\lambda du \\
=& \frac{1}{4\pi} \left[\int_0^1 \frac{1}{\left[g'(\theta_j(u))\right]^2} \times w_k(u) w_l(u)' \left( \intp \frac{1}{f(u,\lambda; \bbeta)} \frac{\partial}{\partial \theta_j(u)} f(u, \lambda; \theta_j(u)) d\lambda \right) du \right]_{k,l},
\end{align*}
for $k,l = 0, \dots, b_j$.
Note that for a fixed $ u $, the inner integral does not depend on rescaled time and can be calculated as in the stationary case. 

\subsection{Inference on parameter curves}
\label{sec:inference:curves}

  In addition to performing statistical inference on the model parameters $ \bbeta $, 
  it is important to infer upon the $J$ time varying parameter curves
  $ \{\theta_j(u) : u \in [0,1]\}, j = 1, \dots, J$.
  Note that 
$\theta_j(u) 
  = \ h_j\!\left(\bd w_j'(u) \bbeta_j\right)$,
  where $ h_j(\cdot) = g_j^{-1}(\cdot) $ is the inverse link function
  for the $j$th curve.
We then use the multivariate delta method to obtain the asymptotic
distribution for the estimated time varying curves at each $u \in [0,1]$,
%
$\widehat{\theta}_{jT}(u) = \ h_j(\bd w_j'(u) \widehat{\bbeta}_{T,j})$,
where $\widehat{\bbeta}_{T,j}$ is the exact ML estimate of
$ \bbeta_j$. The following lemma also holds for the block Whittle
estimator $ \widehat{\bbeta}_{T,j}^W $, but we demonstrate it with
$ \widehat{\bbeta}_{T,j}$ for simplicity.  The following lemma allow
us obtain pointwise confidence bounds for the estimated parameter
curves.

\begin{lemma}\label{asy_normal_curve}
For any $ u \in [0,1] $,
\begin{equation*}
  \sqrt{T}\left(\widehat{\theta}_{T,j}(u) - \theta_{0,j}(u)\right) \xrightarrow{d} \mathcal{N} \left( 0, \nabla \bd H_j \bd w_j'(u) \bd \Gamma(\bbeta_{0,j})^{-1} \bd w_j(u) \nabla \bd H_j \right), \ \quad \text{as} \  T \rightarrow \infty,
  \end{equation*}
  where $ \theta_{0,j}(u) $ is the true $j$th $(j = 1, \dots, J)$ time varying parameter
  curve evaluated at $u$, $ \bd H_j = \mbox{diag} (h_j(\cdot), \ j = 1, \dots, J) $, and $ \{\bd w_j(u)\} $ is a vector of smooth basis functions. 
\end{lemma}



\subsection{Model selection}
\label{sec:model:selection}

In stationary time series, traditional exploratory model selection in
the time domain often involves examining sample autocorrelation and
partial autocorrelation plots of the detrended and deseasonalized
series. Typically, exponentially decaying sample partial
autocorrelations are characteristic of AR models, while similar sample
autocorrelations are indicative of MA models~\cite[see, e.g.,][Section
3.2]{brockwell2002introduction}. Similarly, in the frequency domain,
the periodogram (possibly tapered)~\cite[see, e.g.][Chapter 6]{perc:wald:1993} is used as an estimate of the SDF of the series. Thus, a spike in the periodogram implies that the corresponding frequency is dominant in the time series. These nonparametric approaches are a solid initial step for exploring and identifying possible models. 

The statistical properties of LSB processes vary with time and so we can calculate time varying windowed versions of the above mentioned sample statistics, such as time varying sample autocorrelations or time varying periodograms. These statistics can be used as a tool for carrying out exploratory model selection for LSB processes.  
The time series can be divided into $M$ segments and the statistic of interest (e.g. the sample autocorrelation series or periodogram) is calculated for each of the segments and assigned to the midpoint of the segment. This gives us a rough idea about how the series behaves over time, thus helping in narrowing down the class of models we consider for fitting the LSB process of interest. 
We demonstrate this idea in the EEG example we study in Section~\ref{sec:examples}. 

Model selection for LS processes has been traditionally carried out through various information criteria. \citet{ozaki1975fitting, kitagawa1978procedure, dahlhaus1996maximum, dahlhaus1997fitting} all suggest variations of the Akaike information criterion (AIC) as sufficient model selection criterion, while~\citet{Taniguchi2008} proposes a generalized information criterion based on nonlinear functionals of time varying spectral density, which also reduces to the AIC under certain assumptions. Keeping this in mind, model selection for LSB processes have been done using nonstationary information criteria (NIC)~\citep{dahlhaus1996maximum}, a modified version of AIC, which for the case where the model is correctly specified is
\begin{equation*}
NIC = \mathcal{L}_T(\widehat{\vv{\beta}}_T) + p/T ,
\end{equation*}
where $ \mathcal{L}_T(\widehat{\vv{\beta}}_T) $ is the likelihood function
and $ p $ is the number of parameters that are estimated in the model. 
For a particular LSB model, if we have, say $J$ parameter curves defining that model, then each of those $J$ parameter curves are characterized by $ b_j, \ j = 1, \dots, J $ basis functions. Thus, the model orders $ J, b_1, \dots, b_J $ are chosen using the NIC. 
For example, with an LSB-AR(p) process we have $p+1$
parameter curves, $p$ AR curves and $1$ scale curve, that define the
model. We would calculate the NIC for different values of $p$ and for
different orders of basis functions $ b_1, b_2, \dots, b_{p+1}$ and
select the model that minimizes NIC over the tested grid of $b_j$
values.  To simplify computation, we can carry out this operation in parallel.

\subsection{Forecasting}
\label{sec:forecasting}

Forecasting for time series assuming second order stationarity has been extensively studied in the literature~\citep[e.g.][]{gardner1985exponential, montgomery1990forecasting, box2015time}. The assumption of stationarity implies that the properties of the process remain constant through time, thus allowing for future prediction of the process. The main challenge arises while developing forecasting methodology for nonstationary processes due to their time varying nature. \citet{whittle1965recursive} and \citet{abdrabbo1967prediction} provide some of the early methodology of forecasting procedures for nonstationary time series, while \citet{dahlhaus1996kullback} provides a version of Kolmogorov's formula~\citep[see][Theorem 5.8.1]{brockwell1991time} for forecasting locally stationary time series. In recent years,~\citet{fryzlewicz2003forecasting} provides a forecasting technique for locally stationary wavelet processes, while \citet{palma2013estimation} provides a state-space approach to forecasting locally stationary processes. 

Given observations $ T $ observations $ \{ X_{1,T}, \dots, X_{T,T} \}
$ from an LSB process, suppose we want to obtain a forecast of the $
(T+1)$-th observation $ X_{T+1,T} $.
If we use the past $s$ observations to predict $ X_{T+1,T} $
where $ s = 1, \dots T$,
we have
\begin{align*}
\widehat{X}_{T+1,T} = \sum_{k=1}^{T+1-s} \phi_{T,k}(u)
    X_{T+1-k,T},
\end{align*}
where $\{ \phi_{T,k}(\cdot)$ is calculated using the time varying LD
algorithm outlined in Section \ref{sec:exact}.

%

\section{Testing for departures from stationarity}
\label{sec:test}

Many tests for detecting nonstationarity in time series have been developed in the literature.
\citet{priestley1969test} performed an analysis of variance test using the
log time varying spectral estimates, \citet{von2000wavelet} proposed a
multiple testing procedure based on empirical wavelet coefficients,
and \citet{sakiyama2004discriminant} test for stationarity in a
parametric locally stationary model.  \citet{paparoditis2010validating}
and \citet{dwivedi2011test} also develop spectral based tests. Here,
we construct a likelihood-ratio-type test and derive its asymptotic
distribution under the null hypothesis of weak stationarity.   We
leverage the fact we can parameterize our LSB process to include
stationarity as a special case.

Suppose that $\{ X_{t,T} \}$ is an LSB process, with time varying
transfer function $A(u, \lambda; \vv{\beta}).$  
The process $\{X_{t,T}\} $ is stationary if the transfer function $ A(u, \lambda; \bbeta) = A(\lambda; \bbeta)$ i.e. $ A $ is invariant over rescaled time. This happens when the vector of parameter curves which characterizes the transfer function is constant; i.e., if $ \theta_j(u) = \theta_j,$ for all $ u \in [0,1]$ and for all $ j = 1, \dots, J $. Thus, the parameterization  of the stationary case is nested within that of the LS case. 

Recall from (\ref{basisrep}) that each $ \theta_j(u) $ is modeled as a transformation of linear combination of smooth basis functions given by 
 \begin{align*}
  g_j(\theta_j(u)) = \bd w'_j(u) \bbeta_j 
  = \beta_{j0} + \sum_{l=1}^{b_j} \beta_{jl} w_{jl}(u),    
  \end{align*}
where $ w_{j0}(\cdot) $ describes a constant basis function. Thus, the LSB process simplifies to a stationary process if $ \beta_{jl}$ is zero for $ l = 1, \dots, b_j $ and $ j = 1, \dots, J $. 
 Our hypothesis to test for stationarity can then be defined as 
\begin{align*}
\mathcal{H}_0 :&  \ \beta_{jl} = 0 \  \text{for all} \ l = 1, \dots, b_j \ \text{and} \ j = 1, \dots, J, \mbox{versus}\\
\mathcal{H}_1:& \ \beta_{jl} \neq 0 \ \text{for at least one} \ l =1, \dots, b_j \  \text{and} \  j = 1, \dots, J. 
\end{align*}
Let us partition the model coefficients as $ \bbeta = (\bbeta^{(1)'}, \bbeta^{(2)'})' $ where $ \bbeta^{(1)} = (\beta_{10}, \dots, \beta_{J0})'$ is a $J$ dimensional vector and $ \bbeta^{(2)} = (\beta_{11}, \dots, \beta_{1b_1}, \dots, \beta_{J1}, \dots, \beta_{Jb_j})' $ is a $ (b-J) $ dimensional vector. 
An equivalent test will therefore be
$H_0 : \bbeta^{(2)} = \bd 0$ versus $H_1: \bbeta^{(2)} \neq \bd 0$.
Although the generalized likelihood ratio test proposed here can be constructed using both the likelihood and the block Whittle likelihood, we demonstrate the test for the exact likelihood case given by \eqref{exactlike}.
Let $ \widehat{\bbeta}_T $ is the value of $
\bbeta $ that minimizes the function $ \mathcal{L}_T(\bbeta) $, and
under the null hypothesis, let  $ \tilde{\bbeta}^1_T $ be 
$\tilde{\bbeta}_T^{(1)} = \text{arg}\min_{\bbeta^{(1)}} \mathcal{L}_T(
(\bbeta^{(1)}, \bd 0)' )$.
%
%
%
Then, the generalized likelihood ratio 
test statistic is
%
$ \Lambda = 2T\left\{\mathcal{L}_T( (\tilde{\bbeta}_T^{(1)}, \bd
    0)' ) - \mathcal{L}_T(\widehat{\bbeta})\right\}$.

\begin{theorem}\label{test_dist}
Under $H_0$, $ \Lambda \xrightarrow{d} \chi^2_{b-J}$ as $T \to \infty$. 
\end{theorem}

Thus, we reject the null hypothesis of stationarity when
$\Lambda > \chi^2_{0.95; b-J}$,
where $ \chi^2_{\alpha, \text{df}} $ denotes the $ \alpha $th quantile of a chisquared distribution with df degrees of freedom.  

\section{Simulation studies}
\label{sec:simulation}

\subsection{Parameter estimation}
\label{subsec:verif_large}

In this section, we verify the large sample properties established in
Section~\ref{sec:large:sample} by means of simulation studies for an LSB-AR process. We explain the methodology for simulating such processes and
perform Monte Carlo simulations to compare the large sample properties
of the exact and block Whittle likelihood estimators in these cases.
  
Given a set of $ T $ IID $ \mathcal{N}(0,1) $ random variables and a set of basis functions $ \bd w $ and its corresponding set of basis parameters $ \bbeta $, an LSB-AR process of order $p$ can be simulated in a straightforward manner using the LD algorithm. 
An algorithm to simulating LSB-AR processes of order $p$ is given in
Section S3 of the supplement. We use this algorithm to simulate the LSB-AR processes in the next two illustrations.

\begin{figure}[!t]
\begin{center}
\includegraphics[scale = 1]{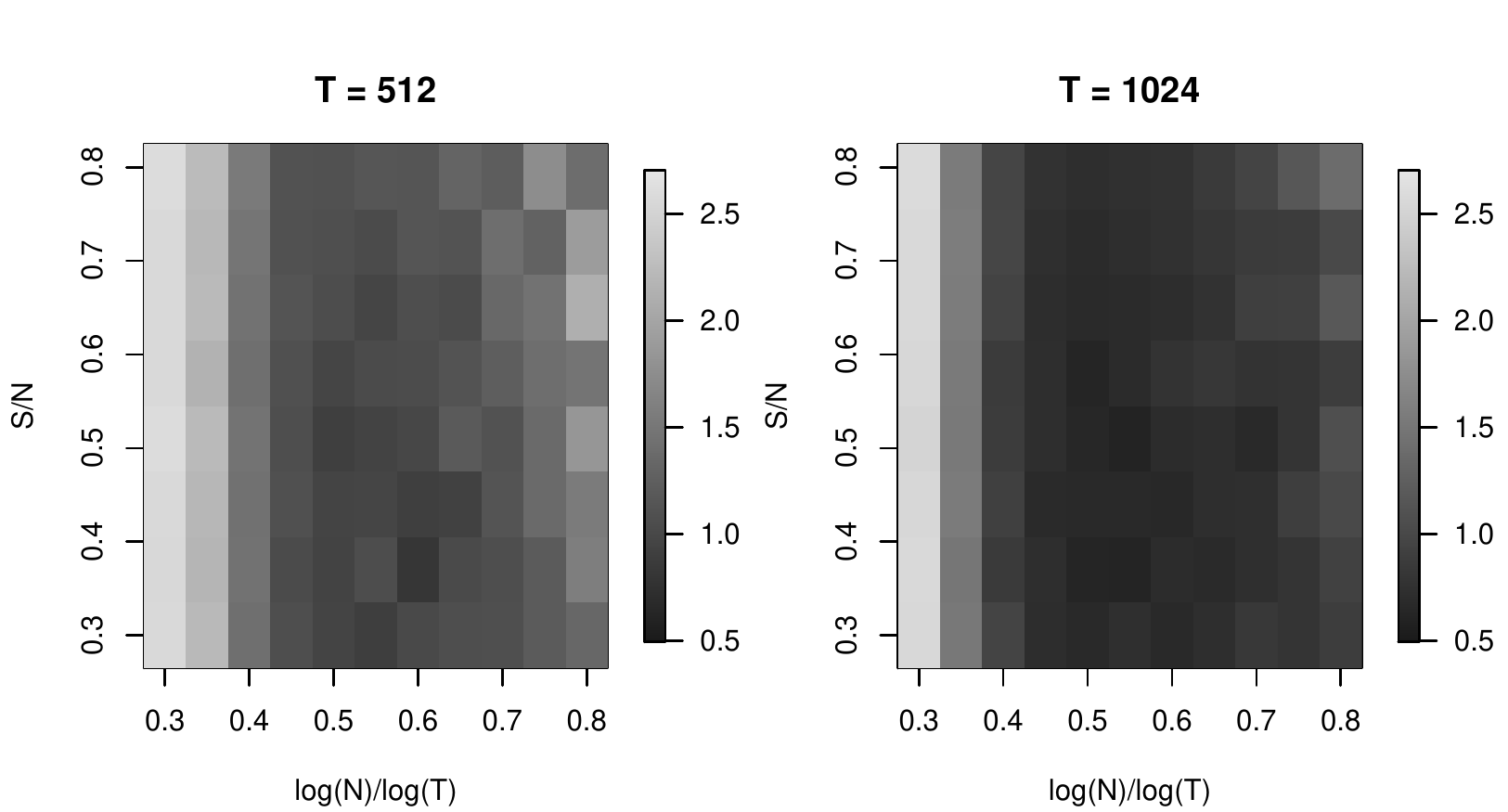} 
\end{center}
\caption{Mean of RMSE($ \widehat{\bbeta}_T^W $) for different values of  $ \log(N)/\log(T)$ and $S/N$.}
\label{NS_selection} 
\end{figure}

\begin{illus}\label{illus:lsb-ar}
In this illustration, we simulate an LSB-AR process of order 2 which slowly transitions to an LSB-AR process of order 1 (see panel (a) of \figurename{ \ref{AR2_FD}} for a plot of the time varying spectral density of this process). We require three basis functions $ \bd w_j(u), j = 1, 2, 3$ to simulate the two time varying partial AR curves and the time varying SD curve. Orthogonal polynomial bases of order 3 are used to simulate the time varying partial AR curves, while a constant basis $ w_3(u) $ is used to simulate the time varying SD curve. The corresponding true parameter vector is $ \bbeta_0 = (\bbeta_{0,1}', \bbeta_{0,2}', \bbeta_{0,3}')' $ where $ \bbeta_{0,1} = (0.61, 1.71, -1.27)'$, $\bbeta_{0,2} = (-3.52, 5.50, -3.00)'$ and $\bbeta_{0,3} = (0)' $. 
We simulate 500 replications and estimate the
model parameter set $ \bbeta_0 $ for each replication using both the
likelihood estimate $ \widehat{\bbeta}_T $ and the block Whittle
likelihood estimate $\widehat{\bbeta}^W_T $ described in
Section~\ref{sec:estimation}. The block Whittle likelihood estimates
are calculated using a cosine bell data taper $ \tau(x) = 0.5[1 -
\cos(2\pi x)]$. We summarize the estimates by taking the mean of all
500 replications. Estimates of bias and RMSE for $ \widehat{\bbeta}_T$
are given by 
\begin{align*}
\text{Bias}\left(\widehat{\bbeta}_T\right) = \frac{1}{500}\sum_{i=1}^{500}\left(\widehat{\bbeta}_T - \bbeta_0\right) \quad \text{and} \quad \text{RMSE}
\left(\widehat{\bbeta}_T\right) = \sqrt{\frac{1}{500}\sum_{i=1}^{500}\left(\widehat{\bbeta}_T - \bbeta_0\right)^2}.
\end{align*}
The bias and RMSE for $ \widehat{\bbeta}_T^W $ are calculated
similarly. Additionally, as a measure of uncertainty, we calculate
$95\%$ bootstrap confidence intervals for the bias and RMSE. We vary
the sample size from $T=128$ to $T=8192$ in powers of two.

As LSB-AR(p) processes are Markov of order $p$, calculating $ \widehat{\bbeta}_T $ is straightforward and efficient, as described in Section~\ref{sec:exact}. It is however interesting to see how much the block Whittle estimate $ \widehat{\bbeta}_T^W $ 
depends on the choice of the block length $N $ and the step size $ S $. To find an appropriate choice of $ N $ and $ S $, we vary the block length $ N $ from $ T^{0.3} $ to $ T^{0.8} $ and the step size $ S $ from $ 0.3 \times N $ to $ 0.8 \times N$. Equivalently, we vary both $\log(N)/\log(T)$ and $ S/N $ from 0.3 to 0.8, calculate the mean of RMSE$\left(\widehat{\bbeta}_T^W\right)$, and see which choice on $N$ and $S$ minimizes this. 
Figure~\ref{NS_selection} provides a plot of these values for sample sizes $ T = 512 $ and $ T = 1024 $. 
As noted from the plot, there is a degree of flexibility in choosing $ N $ and $ S $. For this process, the optimal choice of $ N $ seems to be between $ T^{0.5} $ to $ T^{0.7} $, while the choice of $ S $ seems to be from $ 0.35 \times N $ to $ 0.65 \times N$. The choice of $N$ and $S$ mostly depends on the degree of nonstationarity in the underlying process. Remember that, in the block Whittle estimate, we assume that the data within each block is stationary. So, a larger degree of nonstationarity implies that the stationary assumption will be relatively more plausible for smaller block lengths $N$. In the same way, the optimal step size $S$ will be smaller for a process with a larger degree of nonstationarity, as we would want to lose less information between the blocks.  
In most simulations, we see that the choice of the step size $S$ has less impact on the mean of RMSE$\left(\widehat{\bbeta}_T^W\right) $, compared to the choice of $N$, especially for large sample sizes $T$. This is because as $T$ increases, the information contained in the data increases. For larger sample sizes, increasing the step size loses less information.  

\begin{figure}[t]
\begin{center}
\includegraphics[scale = 1]{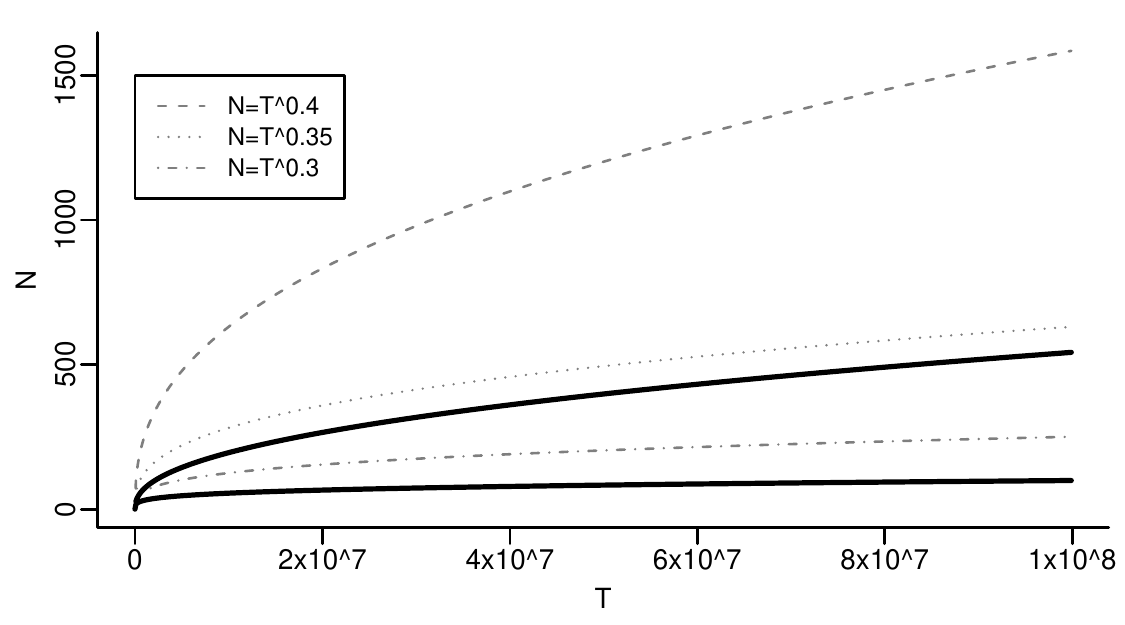} 
\end{center}
\caption{Plot of different choices of the block length $N$ with
increasing sample size $T$. The black solid lines denote the theoretical assumption provided by Dahlhaus.}
\label{NT_plot} 
\end{figure}

The method in which the values of $N$ and $S$ chosen here using Figure~\ref{NS_selection}, are similar to~\citet{palma2010efficient}. 
Although this is an acceptable data adaptive method in choosing $N$ and $S$, it should be noted that, at this time, there is no theoretical method for finding the optimal choice of $N$ and $S$. This is a direction for future research. 
A curious anomaly is that this data adaptive choice on $N$ goes somewhat against Dahlhaus' assumptions given in~\citet{dahlhaus1997fitting} (also given in assumption (AW), required to prove the asymptotic theory. There, the assumption is for $N$ to lie between $ T^{0.25}$ and $ T^{0.5}/\log T $, which, examining Figure~\ref{NT_plot}, is quite a narrow band as seen in Figure~\ref{NT_plot}. However, for practical purposes, choices of $N$ outside this band works well enough for the block Whittle estimator $ \widehat{\bbeta}_T^W $ in terms of minimizing the mean of RMSE$\left(\widehat{\bbeta}_T^W\right) $. 
From now on, $ \widehat{\bbeta}_T^W $ will be calculated using $ N = T^{0.6} $ and $ S = 0.35 N$.

\begin{figure}[p]
\begin{center}
\includegraphics[scale = 0.75]{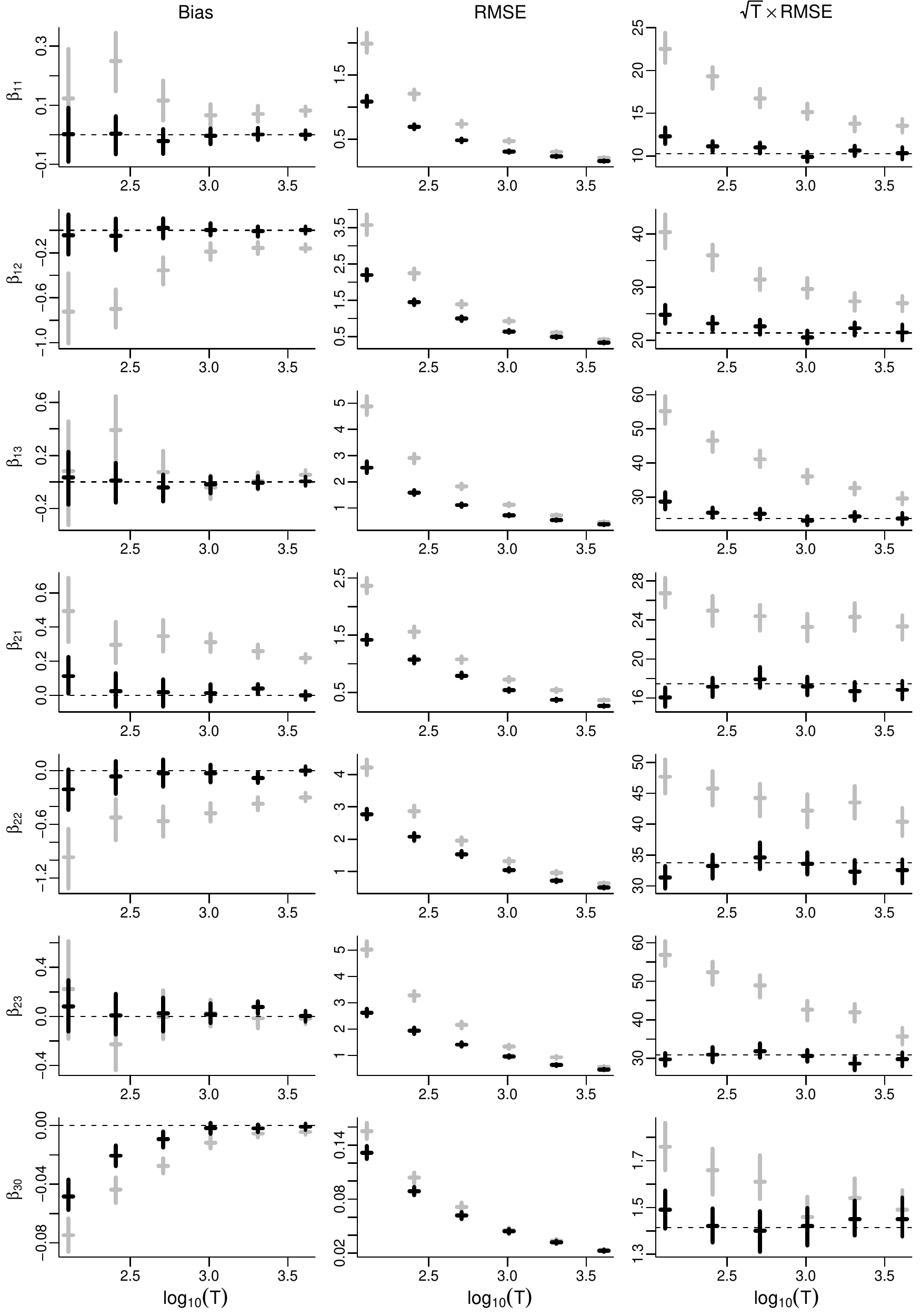}
\end{center}
\caption{Comparison of the Bias, RMSE and $ \sqrt{T\times MSE} $ of the exact (black) and block Whittle (gray) likelihood estimates for increasing log-lengths $\log_{10}T$ of the time series for an LSB-AR process}
\label{tvar2_asymp} 
\end{figure}

\figurename{ \ref{tvar2_asymp}} illustrates a comparison between the estimates of the bias, RMSE and $ \sqrt{T}\times$RMSE for the exact and the Block Whittle likelihood estimates for each element in the parameter vector obtained from the simulation.  
We look at the behavior of these estimates for increasing sample sizes $T$ which serves as an empirical validation for the large sample properties given in Section~\ref{sec:large:sample}. 
The exact likelihood estimates are shown in black with the vertical bar denoting their $95\% $ bootstrapped confidence interval. Similarly, summaries for the block Whittle estimates and associated bootstrap confidence intervals are given in gray.
The first column of Figure~\ref{tvar2_asymp} illustrates that the estimated bias for both $ \widehat{\bbeta}_T $ and $ \widehat{\bbeta}_T^W $ tend to zero (dashed line) with increasing $T$. The estimated bias for $ \widehat{\bbeta}_T $ seems to be smaller than that of $ \widehat{\bbeta}_T^W $. This behavior concurs with the consistency property of the likelihood estimates. The RMSE for the two estimators are shown in the second column of Figure~\ref{tvar2_asymp} and are seen to be decreasing with increasing sample size $T$. The RMSE for the exact likelihood estimator is seen to be consistently smaller than that of the block Whittle estimator.
The third column in Figure~\ref{tvar2_asymp} plots $ \sqrt{T}\times RMSE $ of the two likelihood estimators for increasing sample sizes $T$. From Theorem~\ref{asy normal}, we know that the theoretical asymptotic variance of $ \widehat{\bbeta}_T $ and $ \widehat{\bbeta}_T^W $ is $ \Gamma(\bbeta_0)^{-1} $. 
As $ T $ increases, we expect the bias for the likelihood estimators to tend to zero and so $ \sqrt{T}\times RMSE $ for the estimates of for each parameter $ \bbeta_j, j = 1, \dots, 7 $ should tend to the true $j$th theoretical asymptotic SD $ \Gamma(\bbeta_{0,j})^{-1/2} $ (given by the dashed line). 
This occurs for both estimators as seen in the third column of Figure~\ref{tvar2_asymp} with the confidence intervals around $ \sqrt{T}\times RMSE $ containing the theoretical asymptotic SD for large $T$. 

Although, theoretically, both the estimators have $ \sqrt{T} $ convergence, the exact likelihood estimates seems to be converging to the true asymptotic SD faster than the block Whittle estimates. 
This slower rate of convergence in a practical setting may be due to the fact that we evaluate the Gaussian likelihood by using the Whittle approximation on segments. In each of the $ M $ blocks, we consider the data to be stationary, while, in reality, there is a degree of nonstationarity to the data. Also, in each segment, the Whittle approximation takes advantage of the Toeplitz structure of the covariance matrix to approximate its eigenvalues and eigenvectors. This approximation results in data leakage unless the underlying SDF is constant. As LSB processes have time varying SDF, there is always information leakage, the severity of which depends, again, on the degree of nonstationarity. We try to reduce the bias due to nonstationarity on segments by introducing the cosine bell data taper, which also ensures theoretical $ \sqrt{T}$-consistency of the estimator. However, the speed of convergence still does not match that of the exact likelihood estimator. 
Therefore, for Markov processes such as the LSB-AR process, estimation via exact likelihood might be better due to improved accuracy without significant loss in computational time. 

\end{illus}

\subsection{Testing for departures from stationarity}
\label{verif_power}

We now illustrate the size and power of our test for stationarity that
is described in Section~\ref{sec:test}.  We simulate a LSB-AR($1$)
process of order 1 as defined by \eqref{eq:TVAR},
where the time varying AR parameter $ \phi(\cdot) $ is given by
\begin{align}\label{eq:lsbar1_cond}
\text{logit}\left(\frac{\phi(u)+1}{2}\right) = \beta_{10} + \beta_{11} u,
\end{align}
and we assume $\{  \epsilon_{t,T}\}$ are independent
$\mathcal{N}\left(0,\sigma^2 \right) $ random variables with $ \log(\sigma^2) = \beta_{20} $. 
In this setup, if the coefficient $ \beta_{11} = 0$
in~\eqref{eq:lsbar1_cond}, then the time varying AR curve $ \phi(u) $
is constant over rescaled time $u$, and thus the process $
X_{t,T} $ is stationary. If $ \beta_{11} \neq 0 $, $ X_{t,T} $ is
an LSB-AR(1) process with the degree of nonstationarity depending on
how far the coefficient $ \beta_{11} $ is from zero.  Our test for
stationarity in this case is $H_0 : \beta_{11} = 0$ versus $H_1:
\beta_{11} \neq  0$. 

\begin{figure}[t]
\begin{center}
\includegraphics[height=2.5in]{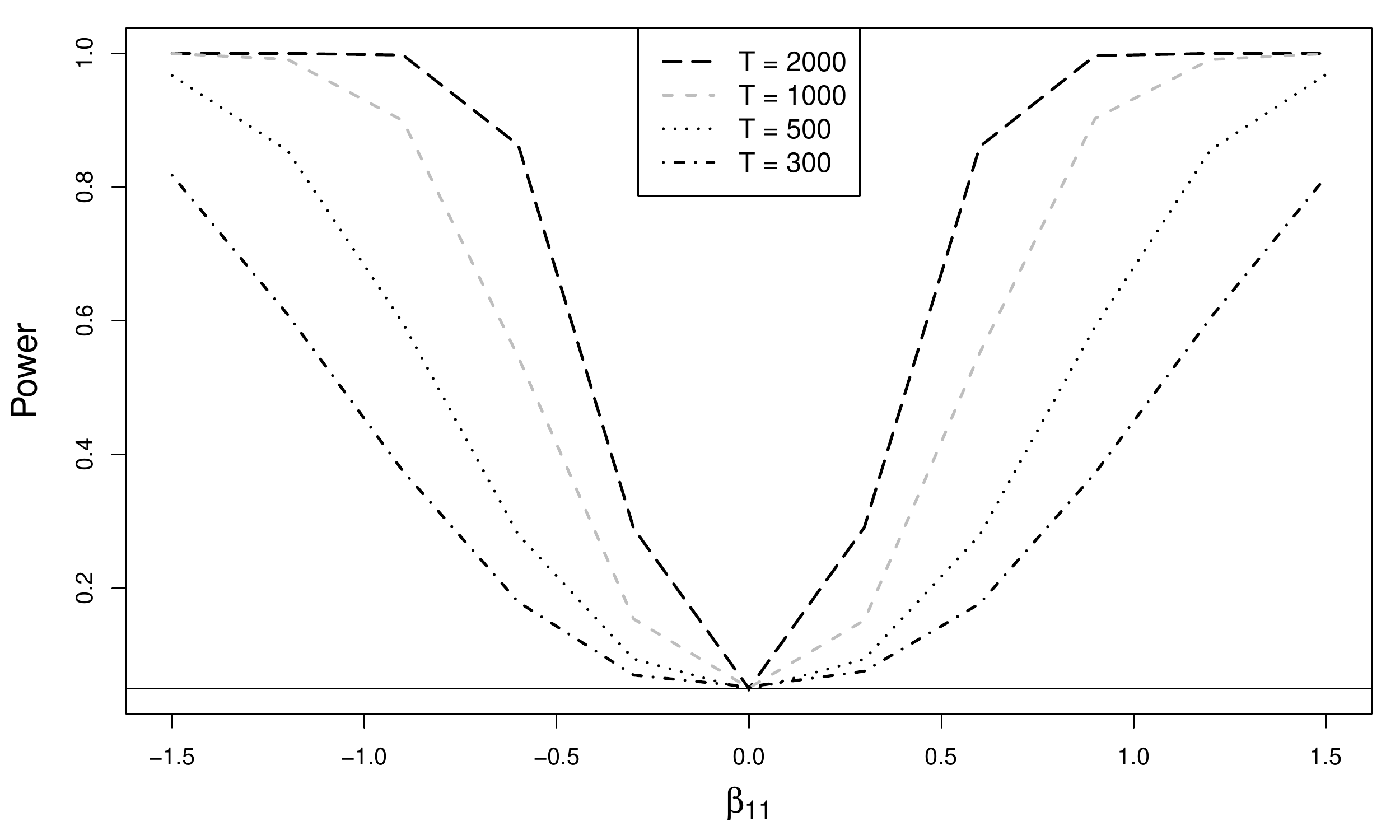} 
\end{center}
\caption{For the LSBAR(1) process, estimated power
  curves as a function of $\beta_{11}$ for different values of $T$.
  The standard errors for each value are no larger than $0.005$.}
\label{AR1_power} 
\end{figure}

For different values of $T$ and $\beta_{11}$, we simulate $10,000$
replications of the LSB-AR(1) process with different values of
$\beta_{11}$.  Table~\ref{tab:size} demonstrates that once we account
for the standard error (which is no greater than $0.005$ in this case)
our test for stationarity contains the nominal level of our test that
was set at $\alpha=0.05$.  Figure~\ref{AR1_power} shows the estimated
power curves as a function of $\beta_{11}$ for different values of
$T$.  As expected the power increases as function of $|\beta_{11}|$,
and the power curves are steeper at longer sample sizes $T$. We
conclude that our test performs as expected even at the smaller sample sizes.

\begin{table}[t]
\centering
\caption{The estimated size of the test for stationarity, as we vary
  the sample size $T$.  The standard errors for each estimated size
  are no larger than $0.005$.}
\begin{tabular}{c|cccc}
Sample size, $T$       & 300             & 500             & 1000            & 2000            \\ \hline
Estimated size     & 0.0527          & 0.0476          & 0.0510           & 0.0499    \\
\end{tabular}
\label{tab:size}
\end{table}

\section{Application to nonstationary EEG Data}
\label{sec:examples}

Electroencephalogram (EEG) time series are collected to non-invasively
monitor electrical activity in the brain.  As the the brain responds
to both internal and external stimuli, we do not expect the
characteristics of brain activity to be constant over time.  Studying how the EEG series
dependence of the electrical signals in the brain change over time
leads naturally to a nonstationary
analysis~\cite[e.g.,][]{kawabata1973nonstationary, ferber1987treatment,
  schiff1994fast, clark1995multiresolution}.
Traditionally, windowed spectral or wavelet analyses are used
as a means to explore these kind of data, however it may be hard to
model and fully account for uncertainty by using these
exploratory tools.  Since we expect signals to change smoothly over
time, we will investigate the use of LS models for the
analysis of EEG series.

\begin{figure}[t]
  \centering
  \includegraphics[height=4in, width=6in]{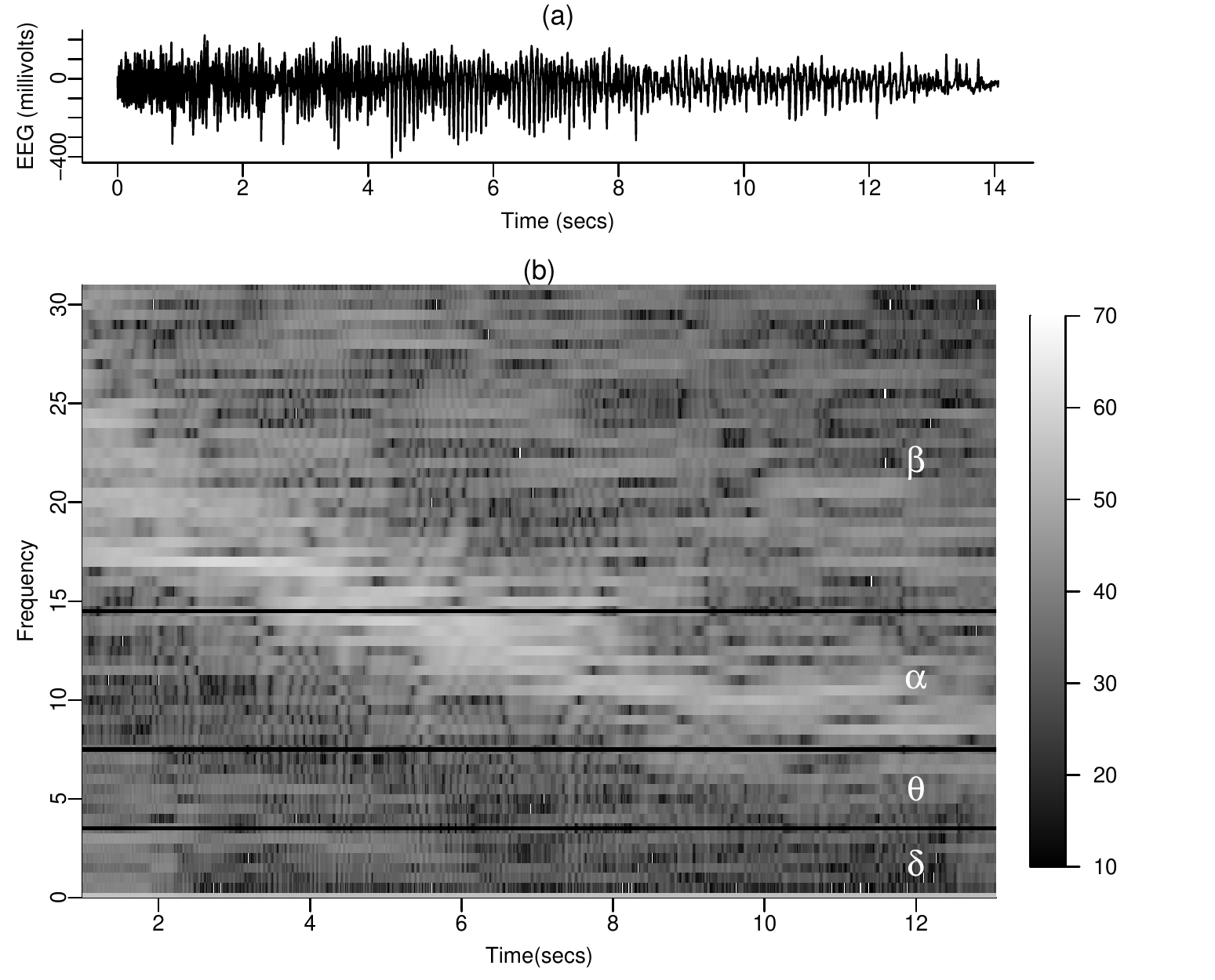}
\caption{(a) EEG series for a patient receiving ECT treatment. (b)
  Windowed estimate of the spectrum with window length = 512.}
\label{fig:EDA} 
\end{figure}

We study the seizure activity in the brain for a subject
undergoing electroconvulsive therapy (ECT), a treatment for patients
with severe clinical depression.  The EEG series comes from
\citet{west1999evaluation}, and is part of an ensemble of 19 series
recorded simultaneously over a patients scalp using Ag/CL
electrodes.  The data is recorded at a sampling rate of 256
observations per second over a period of about 14 seconds.
The EEG series is shown in Figure \ref{fig:EDA}(a) and a windowed
estimate of the SDF is shown in Figure \ref{fig:EDA}(b).  The windowed
estimate uses rolling windows of length 512, and we denote the
different brainwave bands by $\delta$, $\theta$, $\alpha$ and $\beta$
in the plot.  Both the time series plot and
windowed spectral estimate indicate a time varying nature to the process.
We see a dominant spectral peak at around 22 Hz in the
$\beta$ band at the start, and over time the peak moves
to the $\alpha$ band.

Windowed estimates of the partial autocorrelation function (PACF; not shown) indicate that the lag 1 PACF is fairly
constant over time, but that the lag 2 PACF changes over time.  Also, these
PACF plots suggest that there are regions of time for which there are
non-zero PACFs at lags greater than 8 but not greater than lag 20.
Thus we choose to model the series using LSB-AR process of orders
between 8 and 20, as defined in Section~\ref{sec:LSB:examples}.  To
capture the smooth variations in the time varying SDF, we model the
$p$ time varying partial autocorrelation parameter curves
$\{ \phi_{p,j}(u) \}$ and a time varying log SD curve
$\{ \log \sigma(u) \}$ using natural cubic b-splines basis functions,
with equally spaced knots. For each of the $p+1$ curves we use the
same number of basis functions, $b$.

We vary the LSB-AR model order $p$ from 8 to 20 and number of basis
functions $b$ from 2 to 8.  Using the NIC, an LSB-AR(18) process with
$b=4$ basis functions minimizes the criterion.
(An LSB-AR(20) with $b=4$ fits similarly, with respect to
the NIC).  The high order of the time-varying process confirms that
pattern of brain activity is non-trvial and nonstationary.
We formally test whether our series is stationary or not using the
procedure defined in Section~\ref{sec:test}.  A likelihood ratio
statistic of 1404.5 on a chisquared distribution with $60$ degrees of
freedom yields a p-value very close to zero and we reject the null
hypothesis of stationarity for this EEG process.

\begin{figure}[!t]
\begin{center}
\includegraphics[scale = 0.9]{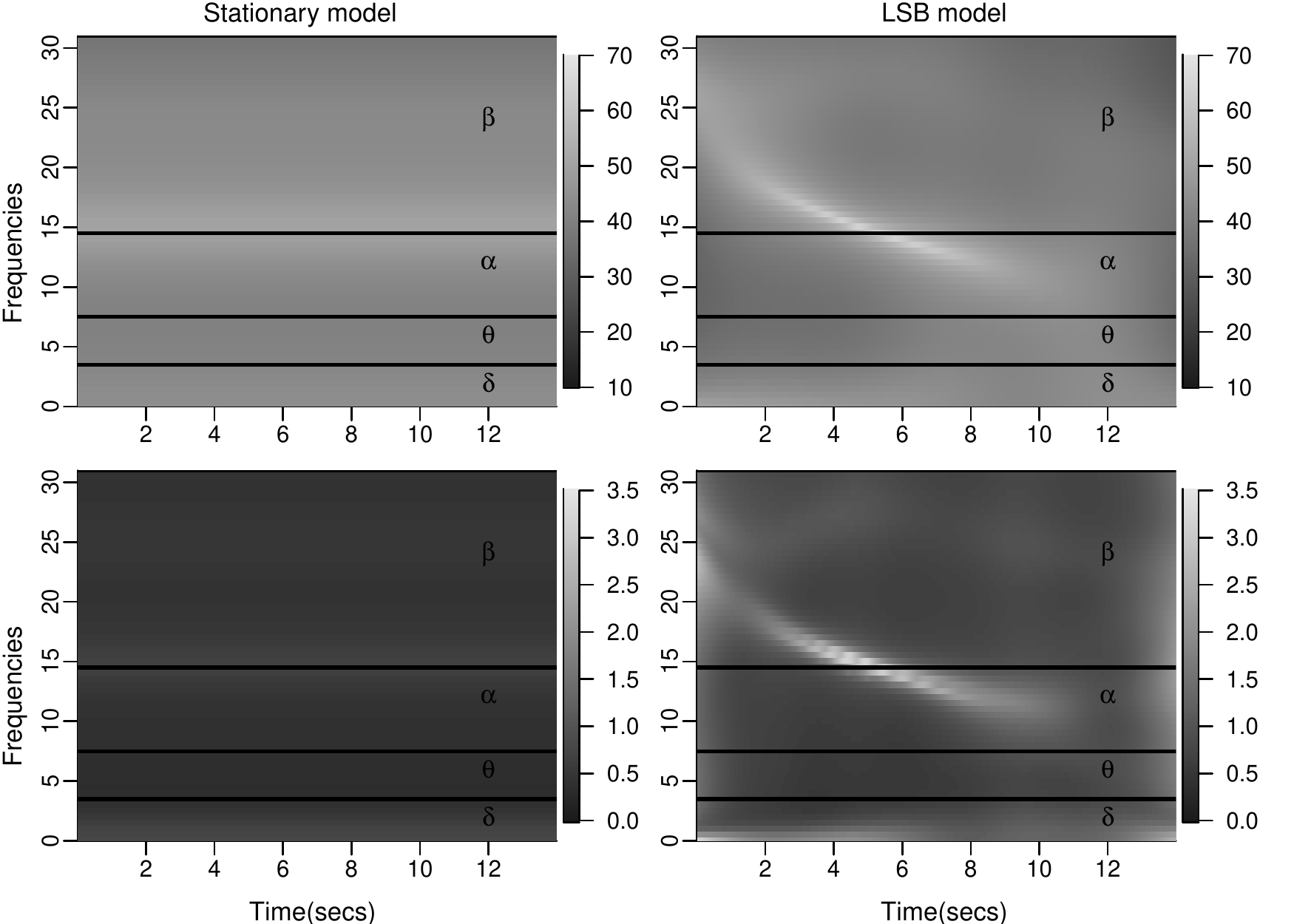} 
\end{center}
\caption{Comparison of the estimate of the time varying
  SDF (top row) and the standard deviation (bottom row) of a stationary AR(18)
  model to an LSB-AR(18) model.}
\label{stat_nonstat_comp} 
\end{figure}

Remember that this EEG series is just one of an ensemble of 19 series
recorded simultaneously over a patient's scalp who is undergoing ECT.
Although, we are analyzing just one location in the scalp, the
inherent nonstationarity in the EEG could be attributed to the ECT
treatment~\citep{krystal1999new}. Figure~\ref{stat_nonstat_comp} gives
a comparison between the estimated, possibly time varying, SDF of this EEG
series using a stationary AR(18) process and our nonstationary
LSB-AR(18) process, along with their corresponding uncertainties.  (In
some cases stationary AR processes have been used to model EEG series;
e.g., \citet{steinberg1985fitting}.)
We confirm that the EEG process for the patient undergoing ECT
originates at around 25 Hz within the beta brainwave band, which
ranges form 15 to 30 Hz. As time evolves, the signal decreases
smoothly and crosses the threshold to the alpha band (8 to 14 Hz) at
around the 5 second mark and keeps on decreasing in frequency. These
smooth variations in the signal are accurately captured by the
LSB-AR(18) process, but not by the stationary process.

\section{Discussion}
\label{sec:conc}

In this article, we introduced the class of LSB processes which are
characterized by time varying parameter curves defined through
transformations of basis functions.  The flexibility of choosing any
continuous basis function means that LSB processes can be used to
model a large class of short and long memory time varying
nonstationary time series processes.  We discussed the statistical
properties of such processes.  An important feature of using LSB
models is that estimation via different likelihood based techniques
are valid and easily implemented for both LSB-SRD and LSB-LRD
processes.  The estimators of the model parameters have good empirical
and theoretical properties.

We have demonstrated likelihood-based modeling and asymptotic results
assuming that the process has zero mean. For LSB processes with a
trend, simultaneous modeling of the time varying mean is important and
can be quite challenging. \citet{dahlhaus1996kullback,
  dahlhaus1996maximum, dahlhaus2000likelihood} and
\citet{dahlhaus2001locally} contain various results for LS processes
with a time varying mean function. Extending results from
\citet{dahlhaus1996maximum}, one can show that for an LSB process
following Definition~\ref{lsb_model} with a time varying mean function
$ \{ \mu(u) = \vv{w}'(u) \bbeta_{\mu}; u \in [0,1]\}, $
Theorem~\ref{asy normal} holds with
\begin{align}\label{inf_mat_mean}
\begin{split}
\bd \Gamma(\bbeta) =&  \frac{1}{4\pi} \int_0^1 \intp (\nabla \log f(u,\lambda; \bbeta))(\nabla \log f(u, \lambda; \bbeta))' d\lambda du \\
& \quad + \frac{1}{2\pi}\int_0^1 \left( \nabla \mu(u) \right)\left( \nabla \mu(u) \right)' f(u, 0; \bbeta)^{-1} du. 
\end{split}
\end{align}

Model selection procedures for LSB processes were discussed using NIC,
an information criteria similar to AIC. While model selection for LS
processes have typically been done through different information
criteria, it would be an interesting problem to develop other types of
model selection methods such as methods based on cross
validation~\citep[e.g.,][]{arlot2010survey} or Bayesian
methods~\citep[e.g.,][]{carlin1995bayesian, dellaportas2002bayesian}.
Related to model selection, we demonstrated in
Section~\ref{sec:simulation} that the choice of block length $N$ and
step size $S$ in nontrivial for the block Whittle likelihood
estimator.  We provided a data adaptive method of choosing these
parameters by minimizing the mean RMSE for these estimators. This
requires a simulation study to be run in practice.  However,
to the best of our knowledge, a theoretical solution to this problem
is yet to be discovered and could be a direction of future research.

The framework for this class of processes can naturally be easily
extended to the class of multivariate time series processes.  Gaussian
likelihood theory for LS processes has already been established
in~\cite{dahlhaus2000likelihood}. We are currently investigating the
extension to multivariate and spatio-temporal LSB processes.

\section*{Acknowledgement}

 Craigmile is supported in part by the US National Science Foundation
 (NSF) under grants DMS-1407604 and SES-1424481, and the National
 Cancer Institute of the National Institutes of Health under grant
 R21CA212308.
We thank Lo-Bin Chang and Christopher Hans for comments that improved this manuscript.


\bibliographystyle{biometrika}
\bibliography{references}

\end{document}